\documentclass[11pt]{article}
\usepackage[english]{babel}
\usepackage[utf8x]{inputenc}
\usepackage{pstricks,theorem,amsmath,amsfonts,amsbsy,amssymb,latexsym}
\usepackage{epsf,psfrag,epsfig,subfigure}
\usepackage{a4wide}
\usepackage{colortbl}
\usepackage{cite}
\usepackage{marvosym,vmargin}
\usepackage{algorithm}
\usepackage[noend]{algorithmic}
\usepackage{boxedminipage,scrtime,ulem}

\newcommand{\new}[1]{\textcolor{blue}{#1}}

\newcommand{\dem}{\noindent \textbf{Proof: }}
\newcommand{\findem}{\hfill $\square~~$\vspace{0.2cm}}

  \renewenvironment{thebibliography}[1]{%
    \begin{oldthebibliography}{#1}%
      \setlength{\parskip}{0ex}%
      \setlength{\itemsep}{0.8ex}%
  }%
  {%
    \end{oldthebibliography}%
  }

\usepackage{tweaklist}
\renewcommand{\enumhook}{\setlength{\topsep}{0pt}%
  \setlength{\itemsep}{0pt}}

\newcommand{\vc}{{\mathbf{vc}}}
\newcommand{\cvc}{{\mathbf{cvc}}}

\newcommand{\w}{{\mathbf{w}}}
\newcommand{\bw}{{\mathbf{bw}}}
\newcommand{\cw}{{\mathbf{cw}}}

\newcommand{\bor}{{\bf bor}}
\newcommand{\mids}{{\bf mid}}
\newcommand{\cuts}{{\bf cut}}

\newcommand{\rep}{{\bf rep}}
\newcommand{\dd}{\mathbf{\mathrm{d}}}

\newcommand{\cupall}{\pmb{\pmb{\cup}}}

\newcommand{\mS}{\mathbb{S}}

\newcommand{\remove}[1]{}

\newtheorem{theorem}[subsection]{Theorem}
\newtheorem{observation}[subsection]{Observation}
\newtheorem{proposition}[subsection]{Proposition}
\newtheorem{corollary}[subsection]{Corollary}
\newtheorem{lemma}[subsection]{Lemma}
\newtheorem{definition}[subsection]{Definition}
\newtheorem{claim}[subsection]{Claim}

\begin{document}

\author{Juanjo Ru\'e\thanks{Laboratorie d'Informatique, \'Ecole Polytechnique, 91128 Palaiseau-Cedex, France. Supported by the European
Research Council under the European Community's 7th Framework Programme,
ERC grant agreement 208471 - ExploreMaps project. E-mail:
       {\tt rue1982@lix.polytechnique.fr}.
         }
         \and Ignasi Sau\thanks{AlGCo project-team, CNRS, Laboratoire d'Informatique, de Robotique et de
         Micro\'electronique de Montpellier (LIRMM), Montpellier, France. Supported by projects ANR Agape and ANR Gratos.
E-mail: \texttt{ignasi.sau@lirmm.fr}.}
\and Dimitrios M. Thilikos\thanks{Department of Mathematics, National and
Kapodistrian University of Athens, Greece. Supported by the project
``Kapodistrias'' (A${\rm \Pi}$ 02839/28.07.2008) of the National and
Kapodistrian University of Athens. E-mail: \texttt{sedthilk@math.uoa.gr}.
}}

\title{Dynamic Programming for Graphs on Surfaces\thanks{The results of this paper were announced in the extended abstract
``\emph{Dynamic Programming for Graphs on Surfaces. Proceedings of
ICALP'2010, volume 6198 of LNCS, pages 372-383}'', which is a combination
of the algorithmic framework presented in this paper and the enumerative
results that can be found in~\cite{RST10_comb_Arxiv}.}}

\date{}

\maketitle
\begin{abstract}\noindent We provide a framework for the design and analysis of
dynamic programming algorithms for surface-embedded graphs on $n$ vertices
and branchwidth at most $k$. Our technique applies to general families of
problems where standard dynamic programming runs in $2^{O(k\cdot \log
k)}\cdot n$ steps.  Our approach combines tools from topological graph
theory and analytic combinatorics. In particular, we introduce a new type
of branch decomposition called {\em surface cut decomposition},
generalizing sphere cut decompositions of planar graphs 
which has nice combinatorial
properties. Namely, the number of partial solutions that can be arranged
on a surface cut decomposition can be upper-bounded by the number of
non-crossing partitions on surfaces with boundary. 
It follows that partial
solutions can be represented by a single-exponential (in the branchwidth
$k$) number of configurations. This proves that, when applied on surface
cut decompositions, dynamic programming runs in $2^{O(k)}\cdot n$ steps.
That way, we considerably extend the class of problems that can be solved
in running times with a {\em single-exponential dependence} on branchwidth
and unify/improve most previous results in this direction.\\

%




\end{abstract}


\section{Introduction}
\label{sec:intro}

One of the most important parameters in the design and analysis of graph
algorithms  is the branchwidth of a graph. Branchwidth, together with its
twin parameter of treewidth, can be seen as a measure of the  topological
resemblance of a graph to a tree. Its algorithmic importance dates back in
the celebrated theorem of Courcelle (see e.g.~\cite{Courcelle89}), stating
that graph problems expressible in Monadic Second Order Logic can be
solved in $f(\bw)\cdot n$ steps (here $\bw$ is the
branchwidth\footnote{The original statement of Courcelle's theorem used
the parameter of treewidth instead of branchwidth. The two parameters are
approximately equivalent, in the sense that one is a constant-factor
approximation of the other.} and $n$ is the number of vertices of the
input graph). Using parameterized complexity terminology, this implies
that a large number of graph problems are fixed-parameter tractable when
parameterized by the branchwidth of their input graph. As the bounds for
$f(\bw)$ provided by Courcelle's theorem are huge, the design of
tailor-made dynamic programming algorithms for specific problems so that
$f(\bw)$ is a simple -- preferably a {\em single-exponential} -- function,
became a natural (and unavoidable) ingredient for many results on graph
algorithms (see~\cite{Arnborg85,Bodlaender88,TelleP97,DFT07}). In this
paper, we provide a general framework for the design and analysis of
dynamic programming algorithms for graphs embedded in surfaces where
$f(\bw)=2^{O(\bw)}$.

\paragraph{Dynamic programming.}
Dynamic programming is applied in a bottom-up fashion on a rooted branch
decomposition the input graph $G$, that roughly is a way to decompose the
graph into a tree structure
 of edge bipartitions (the formal definition is in Section~\ref{sec:prelim}).
Each bipartition defines a separator $S$ of the graph called {\em middle
set}, of cardinality bounded by the branchwidth of the input graph. The
decomposition is routed in the sense that one of the parts of each
bipartition is the ``lower part of the middle set'', i.e., the so-far
processed one. For each graph problem, dynamic programming requires the
suitable definition of tables encoding how potential (global) solutions of
the problem are restricted to a middle set and the corresponding lower
part. The size of these tables reflects the dependence on $k=|S|$ in the
running time of the dynamic programming.

Designing the tables for each middle set $S$ is not always an easy task
and may vary considerably due to the particularities of each problem.
The simplest cases are problems such as {\sc Vertex Cover} and {\sc
Dominating Set}, where the certificate of the solution is a set of
vertices whose choice is not restricted by some global condition. This
directly yields the desired $2^{O(k)}$ upper bound on their size. For
other problems, such as {\sc Longest Path}, {\sc Cycle Packing}, or {\sc
Hamiltonian Cycle}, things are more complicated as the tables encode {\em
pairings of vertices of $S$}, which are $2^{\Theta(k\log k)}$ many.
However, for such problems one can do better for {\sl planar graphs}
following the approach introduced in~\cite{DornPBF10effi}.  The idea
in~\cite{DornPBF10effi} is to use a special type of branch decomposition called
{\em sphere cut decomposition} that can
guarantee that the pairings are \emph{non-crossing} pairings around a
virtual edge-avoiding cycle (called {\em noose}) of the plane where $G$ is
embedded. This restricts the number of tables corresponding to a middle
set $S$ by the $k$-th Catalan number, which is {\em single-exponential} in
$k$.
The same approach was extended for graphs embedded in surfaces of genus
$\gamma$~\cite{DFT06}. The idea in~\cite{DFT06} was to perform a {\sl
planarization} of the input graph by splitting the potential solution into
at most $\gamma$ pieces and then applying the sphere cut decomposition
technique of~\cite{DornPBF10effi} to a more general version of the problem where
the number of pairings is still bounded by some Catalan number
(see also~\cite{DFT08} for the application of  this technique for more
general graphs).

A wider family of problems are those where the tables of dynamic
programming encode {\sl connected packings} of $S$ into sets, i.e.,
collections of subsets of $S$ that are pairwise disjoint and where each
subset is a connected part of a partial solution (see
Section~\ref{sec:expl} for the formal definitions). Throughout this paper,
we call these problems \emph{connected packing-encodable}. Typical
problems of this type are {\sc Connected Vertex Cover}, {\sc Connected
Dominating Set}, {\sc Feedback Vertex Set} ({\sc FVS}), or {\sc Steiner
Tree}, where the connected components of a potential solution can be
encoded by a collection of disjoint subsets of $S$, each of {\sl arbitrary
cardinality}. Here, the general bound on the table size is given by the
$k$-th Bell number, and thus it can again be $2^{\Theta(k\cdot \log k)}$.
To exemplify the differences between distinct types of dynamic programming
encodings, we accompany this paper with an Appendix
are presented (an expert reader may safely skip these examples).
Unfortunately, for the latter category of problems, none of the current
techniques has been able to drop the $2^{\Theta(k\cdot \log k)}$ bound to
a single-exponential one for graphs embedded in surfaces. It is worth
mentioning that, according to the recent lower bounds given by Lokshtanov
\emph{et al}.~\cite{LMS11a}, the bound $2^{\Theta(k\cdot \log k)}$ is best
possible in {\sl general graphs} for some parameterized problems like
\textsc{Disjoint Paths}, unless the Exponential Time Hypothesis (ETH)
fails.

\paragraph{Our results.}
In this paper, we follow a different approach in order to design
single-exponential (in $\bw$) algorithms for graphs embedded in surfaces.
In particular, we deviate  significantly from the planarization technique
of~\cite{DFT06}, which is not able to tackle problems whose solutions are
encoded by general packings. Instead, we extend the concept of sphere cut
decomposition from
 planar graphs to generic surfaces, and we exploit directly the combinatorial structure of the potential solutions in the topological
surface. Our approach permits us to provide in a unified way a
single-exponential (in $\bw$) time analysis for all aforementioned
problems. Examples of other such problems are
\textsc{Maximum Leaf Spanning Tree}, \textsc{Maximum Full-Degree Spanning
Tree},
\textsc{Maximum Leaf Tree}, \textsc{Maximum $d$-Degree-Bounded Connected
Subgraph}, \textsc{Metric TSP}, or \textsc{Maximum $d$-Degree-Bounded
Connected Induced Subgraph} and all the variants studied in~\cite{SaTh10}.
Our results are formally described in Section~\ref{sec:expl} and imply all
the results in~\cite{DFT06,DornPBF10effi}, with running times whose dependence on
genus  is better than the ones in~\cite{DFT06}, as discussed in
Section~\ref{sec:conclusions}.


\paragraph{Our techniques.}
For our results we enhance the current technology of dynamic programming
using, among others, tools from topological graph theory.
Our goal is to define a special type of branch decomposition of embedded
graphs with nice topological properties, which we call \emph{surface cut
decomposition}. Moreover, we prove that such decomposition can be
constructed in single-exponential time.
Surface cut decompositions are based on the concept of {\em polyhedral
decomposition},
which can be constructed in polynomial time.
In the middle sets of a surface cut decomposition,  all vertices, except
possibly a set of cardinality $O(\gamma)$,
 are situated along a set of $O(\gamma)$ nooses of the
surface with $O(\gamma)$ common points. This topological property of the
middle sets
is the source of the single-exponentiality of the size of the tables  in
dynamic programming:  they correspond to non-crossing packings of a set
where all its vertices, except possibly a set of cardinality $O(\gamma)$,
lie on the boundary of a surface.
Our next step is to reduce the problem of counting such packings to the
counting of non-crossing partitions of vertices on the boundary of the
same surface. Then, the single-exponential bound follows by  the recent
enumerative results of~\cite{RST10_comb_Arxiv}.

%


%
For performing dynamic programming, our approach resides in a common
preprocessing step that is to construct a \emph{surface cut
decomposition}.
Then, what remains is just to run a problem-specific dynamic programming
algorithm on such a decomposition. The exponential bound on the size of
the tables of the dynamic programming algorithm follows as a result of the
enumeration analysis in Section~\ref{sec:upperbounds}.

Very recently, a new framework for obtaining {\sl randomized}
single-exponential algorithms parameterized by treewidth in general graphs
has appeared in~\cite{CNP+11}. This framework is based on a dynamic
programming technique named Cut\&Count, which seems applicable to most
connected packing-encodable problems, like {\sc Connected Vertex Cover},
{\sc Connected Dominating Set}, {\sc Feedback Vertex Set}, or {\sc Steiner
Tree}. The randomization in the algorithms of~\cite{CNP+11} comes from the
usage a probabilistic result called the Isolation Lemma~\cite{MVV87},
whose derandomization is a challenging open problem~\cite{ViMu08}.
Therefore, the existence of {\sl deterministic} single-exponential
algorithms parameterized by treewidth for connected packing-encodable
problems in general graphs remains wide open. Our results for graphs on
surfaces, as well as their generalization to any proper minor-free graph
family~\cite{RST11minors}, can be seen as an intermediate step towards an
eventual positive answer to this question.


\paragraph{Organization of the paper.} In
Section~\ref{sec:prelim}, we give the definitions of the main topological
and graph theoretical concepts and  tools that we use in this paper. In
Section~\ref{sec:expl}, we define formally the class of connected
packing-encodable problems and we formally settle the combinatorial
problem of their enumeration. In Section~\ref{sec:poly}, we define the
concept of a polyhedral decomposition. In section~\ref{sec:width}, we give
some results on the behavior of certain width parameters on surfaces and
in Section~\ref{sec:topological_lemmas}, we prove some graph-topological
results.  The concept of a surface-cut decompositions, as well as the
algorithm for its construction, are given in Section~\ref{sec:sphere_cut}.
The enumeration results of the paper are presented in
Section~\ref{sec:upperbounds}. Finally, some conclusions and open problems
are given in Section~\ref{sec:conclusions}.

\section{Preliminaries}
\label{sec:prelim}

\paragraph{Graphs.} We use standard graph terminology, see
for instance~\cite{Diestel05}. All graphs are finite and undirected. Given
a graph $G$ and an edge $e\in E(G)$, let $G\slash e$ be the graph obtained
from $G$ by contracting $e$, removing loops and parallel edges. If $H$ can
be obtained from a subgraph of $G$ by a (possibly empty) sequence of edge
contractions, we say that $H$ is a \emph{minor} of $G$. Given a vertex $u$
with degree two, by \emph{dissolving} $u$ we denote the operation of
replacing $u$ and its two incident edges by an edge between its neighbors.

\paragraph{Topological surfaces.} In this paper, surfaces are compact and their boundary is
homeomorphic to a finite set (possibly empty) of disjoint circles. We
denote by $\beta(\Sigma)$ the number of connected components of the
boundary of a surface $\Sigma$. The Surface Classification
Theorem~\cite{Mohar:graphs-on-surfaces} asserts that a compact and
connected surface without boundary is determined, up to homeomorphism, by
its Euler characteristic $\chi(\Sigma)$ and by whether it is orientable or
not. More precisely, orientable surfaces are obtained by adding $g\geq 0$
\emph{handles} to the sphere $\mS^2$, obtaining the $g$-torus
$\mathbb{T}_g$ with Euler characteristic $\chi(\mathbb{T}_g)=2-2g$, while
non-orientable surfaces are obtained by adding $h > 0$ \emph{cross-caps}
to the sphere, hence obtaining a non-orientable surface $\mathbb{P}_h$
with Euler characteristic $\chi(\mathbb{P}_h) = 2-h$.
A subset $\Pi$ of a surface $\Sigma$ is \emph{surface-separating} if
$\Sigma \setminus \Pi$ has at least two connected components.

As a conclusion, our surfaces are determined, up to homeomorphism, by
their orientability, their Euler characteristic, and the number of
connected components of their boundary. For computational simplicity, it
is convenient to work with the \emph{Euler genus} $\gamma(\Sigma)$ of a
surface $\Sigma$, which is defined as $\gamma(\Sigma)=2-\chi(\Sigma)$.

%

\paragraph{Graphs embedded in surfaces.} Our main reference for graphs on surfaces is the monograph of Mohar and
Thomassen~\cite{Mohar:graphs-on-surfaces}. For a graph $G$ we use the
notation $(G,\tau)$ to denote that $\tau$ is an embedding of $G$ in
$\Sigma$ (that is, a drawing without edge crossings), whenever the surface
$\Sigma$ is clear from the context. An embedding has \emph{vertices},
\emph{edges}, and \emph{faces}, which are zero-, one-, and two-dimensional
open sets, and are denoted $V(G)$, $E(G)$, and $F(G)$, respectively. The
degree $\dd(v)$ of a vertex $v$ is the number of edges incident with $v$,
counted with multiplicity (loops are counted twice).


For a graph $G$, the \emph{Euler genus} of $G$, denoted $\gamma(G)$, is
the smallest Euler genus among all surfaces in which $G$ can be embedded.
Determining the Euler genus of a graph is an \textsc{NP}-hard
problem~\cite{Tho89}, hence we assume throughout the paper that we are
given an already embedded graph.
An \emph{$O$-arc} is a subset of $\Sigma$ homeomorphic to $\mS^1$. A
subset of $\Sigma$ meeting the drawing only at vertices of $G$ is called
\emph{$G$-normal}. If an $O$-arc is $G$-normal, then we call it a
\emph{noose}. The \emph{length} of a noose is the number of its vertices.
Many results in topological graph theory rely on the concept of
\emph{representativity}~\cite{SeymourT94,RS95}, also called
\emph{face-width}, which is a parameter that quantifies local planarity
and density of embeddings. The representativity $\rep(G,\tau)$ of a graph
embedding $(G,\tau)$ is the smallest length of a non-contractible (i.e.,
non null-homotopic) noose in $\Sigma$. We call an embedding $(G,\tau)$
\emph{polyhedral}~\cite{Mohar:graphs-on-surfaces} if $G$ is $3$-connected
and $\rep(G,\tau)\geq 3$, or if $G$ is a clique and $1 \leq |V(G)| \leq
3$. With abuse of notation, we also say in that case that the graph $G$
itself is polyhedral.

For a given embedding $(G,\tau)$, we denote by $(G^*,\tau)$ its dual
embedding. Thus $G^*$ is the geometric dual of $G$. Each vertex $v$ (resp.
face $r$) in $(G,\tau)$ corresponds to some face $v^*$ (resp. vertex
$r^*$) in $(G^*,\tau)$. Also, given a set $S \subseteq E(G)$, we denote by
$S^*$ the set of the duals of the edges in $S$. Let $(G,\tau)$ be an
embedding and let $(G^*,\tau)$ be its dual. We define the \emph{radial
graph embedding} $(R_G,\tau)$ of $(G,\tau)$ (also known as
\emph{vertex-face graph embedding}) as follows: $R_G$ is an embedded
bipartite graph with vertex set $V(R_G)= V(G) \cup V(G^*)$. For each pair
$e=\{v,u\}$, $e^*=\{u^*,v^*\}$ of dual edges in $G$ and $G^*$, $R_G$
contains edges $\{v,v^*\}$, $\{v^*,u\}$, $\{u,u^*\}$, and $\{u^*,v\}$.
Mohar and Thomassen~\cite{Mohar:graphs-on-surfaces} proved that, if
$|V(G)|\geq 4$, the following conditions are equivalent: \emph{(i)}
$(G,\tau)$ is a polyhedral embedding; \emph{(ii)} $(G^*,\tau)$ is a
polyhedral embedding; and \emph{(iii)} $(R_G,\tau)$ has no multiple edges
and every 4-cycle of $R_G$ is the border of some face. The \emph{medial
graph embedding} $(M_G,\tau)$ of $(G,\tau)$ is the dual embedding of the
radial embedding $(R_G,\tau)$ of $(G,\tau)$. Note that $(M_G,\tau)$ is a
$\Sigma$-embedded 4-regular graph. \vspace{-2mm}

\paragraph{Tree-like decompositions of graphs.}  Let $G$ be a graph on $n$
vertices. A {\em branch decomposition}~$(T,\mu)$ of a graph $G$ consists
of an unrooted ternary tree $T$ (i.e., all internal vertices are of degree
three) and a bijection $\mu: L \rightarrow E(G)$ from the set $L$ of
leaves of $T$ to the edge set of $G$. We define for every edge~$e$ of~$T$
the {\em middle set}~$\mids(e) \subseteq V(G)$ as follows: Let $T_1$ and
$T_2$ be the two connected components of $T\setminus \{e\}$. Then let
$G_i$ be the graph induced by the edge~set~$\{\mu(f): f \in L \cap V(T_i)
\}$ for $i \in \{1,2\}$. The \emph{middle set} is the intersection of the
vertex sets of $G_1$ and $G_2$, i.e., $\mids(e):= V(G_1) \cap V(G_2)$. The
{\em width} of $(T,\mu)$ is the maximum order of the middle sets over all
edges of $T$, i.e., $\w(T,\mu) := \max\{|\mids(e)| \mid e\in T\}$. An
optimal branch decomposition of $G$ is defined by a tree $T$ and a
bijection $\mu$ which give the minimum width, the {\em branchwidth},
denoted by $\bw(G)$.

Let $G=(V,E)$ be a connected graph. For $S \subseteq V$, we denote by
$\delta(S)$ the set of all edges with an end in $S$ and an end in $V
\setminus S$. Let $\{V_1,V_2\}$ be a partition of $V$. If $G[V \setminus
V_1]$ and $G[V \setminus V_2]$ are both non-null and connected, we call
$\delta(V_1)$ a \emph{bond} of $G$~\cite{SeymourT94}.

 A \emph{carving decomposition}~$(T,\mu)$ is similar to a branch
decomposition, only with the difference that $\mu$ is a bijection between
the leaves of the tree and the vertex set of the graph $G$. For an edge
$e$ of $T$, the counterpart of the middle set, called the \emph{cut set}
$\cuts(e)$, contains the edges of $G$ with endvertices in the leaves of
both subtrees. The counterpart of branchwidth is \emph{carvingwidth}, and
is denoted by $\cw(G)$. In a \emph{bond carving decomposition}, every cut
set is a bond of the graph. That is, in a bond carving decomposition,
every cut set separates the graph into two connected components.

Let $G_1$ and $G_2$ be graphs with disjoint vertex-sets and let $k \geq 0$
be an integer. For $i=1,2$, let $W_i \subseteq V(G_i)$ form a clique of
size $k$ and let $G_i'$ ($i=1,2$) be obtained from $G_i$ by deleting some
(possibly no) edges from $G_i[W_i]$ with both endvertices in $W_i$.
Consider a bijection $h: W_1 \to W_2$. We define a \emph{clique sum} $G$
of $G_1$ and $G_2$, denoted by $G=G_1\oplus_k G_2$, to be the graph
obtained from the union of $G'_1$ and $G_2'$ by identifying $w$ with
$h(w)$ for all $w \in W_1$. The integer $k$ is called the \emph{size} of
the clique sum. Given a set of graphs $\mathcal{G}$ and an integer $\ell
\geq 0$, we define the \emph{$\ell$-clique sum closure} of $\mathcal{G}$
as the set of graphs $\mathcal{G}_{\ell}$ recursively defined as follows:
every graph in $\mathcal{G}$ is also in  $\mathcal{G}_{\ell}$, and if $G_1
\in \mathcal{G}$, $G_2 \in \mathcal{G}_{\ell}$, and $G_3 = G_1 \oplus_k
G_2$ with $0 \leq k \leq \ell$, then $G_3 \in \mathcal{G}_{\ell}$.


\section{Connected packing-encodable problems}
\label{sec:expl}

The standard dynamic programming approach on branch decompositions
requires the so called \emph{rooted} branch decomposition, defined as a
triple $(T,\mu,e_{r})$, where $(T,\mu)$ is a branch-decomposition of $G$
such that $T$ is a tree rooted on a leaf $v_{l}$ of $T$ incident with some
edge $e_{r}$. We slightly abuse notation by insisting that no edge of $G$
is assigned to $v_{l}$ and thus $\mids(e_{r})=\emptyset$ (for this, we
arbitrarily pick some edge of a branch decomposition, subdivide it and
then connect by $e_{r}$ the subdivision vertex with a new leaf $v_{l}$).
The edges of $T$ are oriented towards the root $e_{r}$ and for each edge
$e\in E(T)$ we denote by $E_{e}$ the edges of $G$ that are mapped to
leaves of $T$ that are descendants of $e$. We also set $G_{e}=G[E_{e}]$
and we denote by $L(T)$ the edges of $T$ that are incident with leaves of
$T$. Given an edge $e$ whose tail is a non-leaf vertex $v$, we denote by
$e_1,e_2 \in E(T)$ the two edges heading at $v$ (we call them {\em
children} of $e$). When the tail of an edge of $T$ is also a leaf of $T$
then we call it \emph{leaf-edge}.

Typically, dynamic programming on a rooted branch decomposition
$(T,\mu,e_{r})$ of a graph $G$ associates some suitable combinatorial
structure ${\sf struct}(e)$ with each edge
\new{$e$} of $T$ such that the knowledge of ${\sf struct}(e_{r})$
makes it possible to determine the solution to the problem. Roughly
speaking, ${\sf struct}(e)$ encodes all the ways that the possible
certificates of a partial solution on graph $G_{e}$ may be restricted to
$\mids(e)$. The computation of ${\sf struct}(e)$ is done bottom-up by
first providing ${\sf struct}(e)$ when $e$ is a leaf-edge of $T$ and then
giving a recursive way to construct ${\sf struct}(e)$ from ${\sf
struct}(e_{1})$ and ${\sf struct}(e_{2})$, where $e_{1}$ and $e_{2}$ are
the children of $e$.

The encoding of ${\sf struct}$ is commonly referred as  the ``tables'' of
the dynamic programming algorithm. It is desirable that the size of the
tables, as well as the time to process them, is bounded by
$f(|\mids(e)|)\cdot n^{O(1)}$, where $f$ is a function not depending on
$n$. This would give a polynomial-time algorithm for graphs of fixed
branchwidth. In technical terms, this means that the problem is {\sl Fixed
Parameter Tractable} {\sf(FPT}), when parameterized by the branchwidth of
the input graph (for more on Fixed Parameter Tractability,
see~\cite{FlGr06,DoFe99,Niedermeier06inv}).
%
%
%
A challenge in the design of such algorithms is to reduce the contribution
of branchwidth to the size of their tables and therefore to simplify $f$
as much as possible. As indicated by the lower bounds
in~\cite{LMS11b,ImpagliazzoPZ01whi,CaiJ03onth}, for many problems like
\textsc{Independent Set}, \textsc{Dominating Set}, or
$q$-\textsc{Coloring} for fixed $q \geq 3$, $f$ is not expected to be
better than single-exponential in general graphs.

Before we proceed with the description of the family of problems that we
examine in this paper, we need some definitions. Let $G$ be a graph and
let $S$ be a set of vertices of $G$. We denote by ${\cal G}$ the
collection of all subgraphs of $G$. Each $H\in {\cal G}$ defines a packing
${\cal P}_{S}(H)$ of $S$ such that two vertices $x,y\in S$ belong to the
same set of ${\cal P}_{S}(H)$ if $x,y$ belong to the same connected
component of $H$. We say that $H_{1},H_{2}\in {\cal G}$ are {\em
$S$-equivalent} if ${\cal P}_{S}(H_{1})={\cal P}_{S}(H_{2})$, and we
denote it by $H_{1}\equiv_{S} H_{2}$. Let $\overline{{\cal G}}_{S}$ the
collection of all subgraphs of $G$ modulo the equivalence relation
$\equiv_{S}$. We define the set of all {\em connected packings of $S$ with
respect to $G$} as the collection
$$\Psi_{G}(S)=\{{\cal P}_{S}(H)\mid H \in \overline{{\cal G}}_{S}\}.$$
%
%
%
%
Notice that each member of $\Psi_{G}(S)$ can indeed be seen as a packing
of $S$, as its sets may not necessarily meet all vertices of $S$.
\medskip

In this paper we consider
 graph problems
that can be solved by dynamic programming algorithms on branch
decompositions for  which the  size of ${\sf struct}(e)$ is upper-bounded
by $2^{O(|\mids(e)|)}\cdot |\Psi_{G_{e}}(\mids(e))|\cdot n^{O(1)}$.  We
call these problems {\em connected packing-encodable}. We stress that our
definition of connected packing-encodable problem assumes the existence of
{\sl an} algorithm with this property, but there may exist other
algorithms whose tables are much bigger. In the introduction, we gave a
long list of problems that belong to this category and, in the Appendix,
we make a full description on how to do dynamic programming for one of
them. For these problems, dynamic programming has a single-exponential
dependance on branchwidth if and only if $\Psi_{G_{e}}(\mids(e))$ contains
a single-exponential number of packings, i.e.,
$|\Psi_{G_{e}}(\mids(e))|=2^{O(|\mids(e)|)}$.

However, in general the number of different connected packings that could
be created during the dynamic programming is not necessarily smaller than
the number of the non-connected ones. Therefore, it may linearly depend on
the $k$-th Bell number, where $k$ is the branchwidth of the input graph.
This implies that, in general, $|\Psi_{G_{e}}(\mids(e))|=2^{O(k\log k)}$
is the best upper bound we may achieve for connected packing-encodable
problems, at least for deterministic algorithms. The purpose of this paper
is to show that, for such problems, this bound can be reduced to a
single-exponential one when their input graphs have bounded genus. In
Section~\ref{sec:sphere_cut}, we define the concept of a surface cut
decomposition, which is a key tool for the main result of this paper,
resumed as follows.

\begin{theorem}
\label{teo:fidnal} Every connected packing-encodable problem whose input
graph $G$  is embedded in a surface of Euler genus $\gamma$, and has
branchwidth at most $k$, can be solved by a dynamic programming algorithm
on a surface cut decomposition of $G$ with tables of size $\gamma^{O(k)}
\cdot k^{O(\gamma)}\cdot \gamma^{O(\gamma)}\cdot n^{O(1)}$.
\end{theorem}

In Section~\ref{sec:sphere_cut}, we prove (Theorem~\ref{teo:surface_cut})
that, given a graph $G$ embedded in a surface of Euler genus $\gamma$, a
surface cut decomposition of $G$ of width $O(\bw(G)+\gamma)$ can be
constructed in $2^{O(\bw(G))}\cdot n^{3}$ steps. Therefore, we conclude
the following result.

\begin{theorem}
\label{teo:finals} Every connected packing-encodable problem whose input
graph $G$  is embedded in a surface of Euler genus $\gamma$, and has
branchwidth at most $k$, can be solved in $\gamma^{O(k)} \cdot
k^{O(\gamma)} \cdot \gamma^{O(\gamma)}\cdot n^{O(1)}$ steps.
\end{theorem}

Given a parameterized problem with parameter $k$, an algorithm that solves
it in time $2^{O(k)}\cdot n^{O(1)}$ is called {\em single-exponential {\sf
FPT}-algorithm}. As finding an optimal embedding of a graph of genus
$\gamma$ can be solved in $f(\gamma)\cdot n$ steps~\cite{Mohar99alin}, we
can restate Theorem~\ref{teo:finals} as follows.

\begin{corollary}
Every connected packing-encodable problem on graphs of fixed genus has a
single-exponential {\sf FPT}-algorithm, when parameterized by the
branchwidth of its input.
\end{corollary}

\section{Polyhedral decompositions}
\label{sec:poly}  We introduce in this section \emph{polyhedral
decompositions} of graphs embedded in surfaces.
Let $G$ be an embedded graph, and let $N$ be a noose in the surface.
Similarly to~\cite{CaMo07}, we use the notation $G$\Leftscissors$N$ for
the graph obtained by cutting $G$ along the noose $N$ and gluing a disk on
the obtained boundaries.


\begin{definition}
\label{def:poly} Given a graph $G=(V,E)$ embedded in a surface of Euler
genus $\gamma$, a \emph{polyhedral decomposition} of $G$ is a set of
graphs $\mathcal{G}=\{H_1,\ldots,H_{\ell}\}$ together with a set of
vertices $A \subseteq V$ such that
\begin{itemize}
\item $|A|=O(\gamma)$;\vspace{-0.12cm}
\item $H_i$ is a minor of $G[V\setminus A]$, for $i=1,\ldots,\ell$;\vspace{-0.12cm}
\item $H_i$ has a polyhedral embedding in a surface of Euler genus at most $\gamma$, for $i=1,\ldots,\ell$;\vspace{-0.12cm}
\item $G[V\setminus A]$ belongs to the $2$-clique sum closure of $\mathcal{G}$.
\end{itemize}
\end{definition}
\vspace{-0.4cm}


\begin{observation}
\label{obs:noose} Note that an embedded graph $H$ is not polyhedral if and
only if there exists a noose $N$ of length at most two in the surface in
which $H$ is embedded, such that either $N$ is non-contractible or $V(H)
\cap N$ separates $H$. Indeed, if $H$ has representativity at most two,
then there exists a non-contractible noose $N$ of length at most two.
Otherwise, since $H$ is not polyhedral, $H$ has a minimal separator $S$ of
size at most two. It is then easy to see that there exists a noose
containing only vertices of $S$.
\end{observation}

Algorithm~\ref{alg:polyhedral} provides an efficient way to construct a
polyhedral decomposition, as it is stated in Proposition~\ref{prop:poly}.
In the algorithm, the addition of an edge $\{u,v\}$ represents the
existence of a path in $G$ between $u$ and $v$ that is not contained in
the current component.


\begin{algorithm}[htb]
{\small
\caption{Construction of a polyhedral decomposition of an embedded graph
$G$} \label{alg:polyhedral}
\begin{algorithmic} 
\REQUIRE{A graph $G$ embedded in a surface of Euler genus $\gamma$.}
\ENSURE{A polyhedral decomposition of $G$.}
\medskip
\STATE{$A=\emptyset$, $\mathcal{G} = \{G\}$ (the elements in
$\mathcal{G}$, which are embedded graphs, are called \emph{components}).}

\WHILE{$\mathcal{G}$ contains a non-polyhedral component $H$}

\STATE{Let $N$ be a noose as described in Observation~\ref{obs:noose} in
the surface in which $H$ is embedded,\\ and let $S=V(H) \cap N$.}

\IF{$N$ is non-surface-separating}

    \STATE{Add $S$ to $A$, and replace in $\mathcal{G}$ component $H$ with $H[V(H)\setminus
    S]$\Leftscissors$N$.}

    \ENDIF

\IF{$N$ is surface-separating}

    \STATE{Let $H_1$, $H_2$ be the subgraphs of $H$\Leftscissors$N$ corresponding to the two surfaces occurring after splitting $H$}

    \IF{$S=\{u\} \cup \{v\}$ and $\{u,v\} \notin E(H)$}
        \STATE{Add the edge $\{u,v\}$ to $H_i$, $i=1,2$.}
    \ENDIF

    \STATE{Replace in $\mathcal{G}$ component $H$
         with the components of $H$\Leftscissors$N$ containing at least one edge of $H$.}
\ENDIF
%
\ENDWHILE \RETURN{$(\mathcal{G},A)$.}
\end{algorithmic}
}
\end{algorithm}


\begin{proposition}
\label{prop:poly} Given a graph $G$ on $n$ vertices embedded in a surface,
Algorithm~\ref{alg:polyhedral} constructs a polyhedral decomposition of
$G$ in $O(n^3)$ steps.
\end{proposition}
\dem We first prove that the the output $(\mathcal{G},A)$ of
Algorithm~\ref{alg:polyhedral} is indeed a polyhedral decomposition of
$G$, and then we analyze the running time.

Let us see that each component of $\mathcal{G}$ is a minor of
$G[V\setminus A]$. Indeed, the only edges added to $G$ by
Algorithm~\ref{alg:polyhedral} are those between two non-adjacent vertices
$u,v$ that separate a component $H$ into several components
$H_1,\ldots,H_{\ell}$. For each component $H_i$, $i=1,\ldots,\ell$, there
exists a path between $u$ and $v$ in $H \setminus H_i$ (provided that the
separators of size 1 have been already removed, which can we assumed
without loss of generality), and therefore the graph obtained from $H_i$
by adding the edge $\{u,v\}$ is a minor of $H$, which is inductively a
minor of $G[V\setminus A]$. Also, each component of $\mathcal{G}$ is
polyhedral by definition of the algorithm.

As a non-separating noose is necessarily non-contractible, each time some
vertices are moved to $A$, the Euler genus of the surfaces strictly
decreases~\cite[Lemma 4.2.4]{Mohar:graphs-on-surfaces}. Therefore,
$|A|=O(\gamma)$.

By the construction of the algorithm, it is also clear that each component
of $\mathcal{G}$ has a polyhedral embedding in a surface of Euler genus at
most $\gamma$. Finally, $G[V\setminus A]$ can be constructed by joining
the graphs of $\mathcal{G}$ applying clique sums of size at most two.

Thus, $(\mathcal{G},A)$ is a polyhedral decomposition of $G$ according to
Definition~\ref{def:poly}.


We now analyze the running time of the algorithm. Separators of size at
most two can be found in $O(n^2)$ steps~\cite{HRG00}. A noose with respect
to a graph $H$ corresponds to a cycle in the radial graph of $H$, hence
can also be found\footnote{A shortest non-contractible cycle can be found
in $2^{O(\gamma \log \gamma)}n^{4/3}$ steps~\cite{CaMo07}. This running
time improves on $O(n^3)$ for a big range of values of $\gamma$.} in
$O(n^2)$ (using that the number of edges of a bounded-genus graph is
linearly bounded by its number of vertices). Since each time that we find
a {\sl small} separator we decrease the size of the components, the
running time of the algorithm is $O(n^3)$. \findem

\section{Width parameters of graphs on surfaces}
\label{sec:width}

In this section we state some definitions and auxiliary results about
several width parameters of graphs on surfaces, to be applied in
Section~\ref{sec:sphere_cut} for building surface cut decompositions. In
the same spirit of~\cite[Theorem 1]{FoTh07} we can prove the following
lemma. We omit the proof here since the details are very
similar\footnote{The improvement in the multiplicative factor of the Euler
genus is obtained by applying more carefully Euler's formula in the proof
analogous to that of~\cite[Lemma 2]{FoTh07}.} to the proof
in~\cite{FoTh07}.
\begin{lemma}
\label{lem:carving12} Let $(G,\tau)$ and $(G^*,\tau)$ be dual polyhedral
embeddings in a surface of Euler genus $\gamma$ and let $(M_G,\tau)$ be
the medial graph embedding. Then $\max\{\bw(G),\bw(G^*)\} \leq \cw(M_G)/2
\leq 6 \cdot \bw(G) + 2\gamma + O(1).$ In addition, given a branch
decomposition of $G$ of width at most $ k$, a carving decomposition of
$M_G$ of width at most $12k$ can be found in linear time.
\end{lemma}

We would like to point out that in Lemma~\ref{lem:carving12} we need the
embeddings to be polyhedral.

\begin{lemma}[folklore]
\label{lem:bw1} The removal of a vertex from a non-acyclic graph decreases
its branchwidth by at most $1$.
\end{lemma}
\begin{lemma}
\label{lem:bw2} Let $G$ be a graph and let $\mathcal{G}$ be a collection
of graphs such that $G$ can be constructed by joining graphs in
$\mathcal{G}$ applying clique sums of size at most two. Given branch
decompositions $\{(T_H,\mu_H)\ |\ H \in \mathcal{G})\}$, we can compute in
linear time a branch decomposition $(T,\mu)$ of $G$ such that $\w(T,\mu)
\leq \max \{ 2, \{ \w(T_H,\mu_H)\ |\ H \in \mathcal{G}\} \}$. In
particular, $\bw(G) \leq \max \{ 2, \{ \bw(H)\ |\ H \in \mathcal{G}\} \}$.
\end{lemma}
\dem Note that if $G_1$ and $G_2$ are graphs with no vertex (resp. a
vertex, an edge) in common, then $G_1 \cup G_2 = G_1 \oplus_0 G_2$ (resp.
$G_1 \oplus_1 G_2$, $G_1 \oplus_2 G_2$). To prove Lemma~\ref{lem:bw2}, we
need the following two lemmata.
\begin{lemma}
\label{lem:sum1} Let $G_1$ and $G_2$ be graphs with at most one vertex in
common. Then $\bw(G_1 \cup G_2) = \max \{\bw(G_1),\bw(G_2)\}$.
\end{lemma}
\dem Assume first that $G_1$ and $G_2$ share one vertex $v$. Clearly
$\bw(G_1 \cup G_2) \geq \max \{\bw(G_1),\bw(G_2)\}$. Conversely, for
$i=1,2$, let $(T_i,\mu_i)$ be a branch decomposition of $G_i$ such that
$\w(T_i,\mu_i)\leq k$. For $i=1,2$, let $T_i^v$ be the minimal subtree of
$T_i$ containing all the leaves $u_i$ of $T_i$ such that $v$ is an
endvertex of $\mu_i(u_i)$. For $i=1,2$, we take an arbitrary edge
$\{a_i,b_i\}$ of $T_i^v$, we subdivide it by adding a new vertex $w_i$,
and then we build a tree $T$ from $T_1$ and $T_2$ by adding the edge
$\{w_1,w_2\}$. We claim that $(T,\mu_1 \cup \mu_2)$ is a branch
decomposition of $G_1 \cup G_2$ of width at most $k$. Indeed, let us
compare the middle sets of $(T,\mu_1 \cup \mu_2)$ to those of
$(T_1,\mu_1)$ and $(T_2,\mu_2)$. First, it is clear that the vertices of
$V(G_1) \cup V(G_2) - \{v\}$ appear in $(T,\mu_1 \cup \mu_2)$ in the same
middle sets as in $(T_1,\mu_1)$ and $(T_2,\mu_2)$. Secondly,
$\mids(\{w_1,w_2\})=\{v\}$, since $v$ is a cut-vertex of $G_1 \cup G_2$.
Also, for $i=1,2$,
$\mids(\{a_i,w_i\})=\mids(\{w_i,b_i\})=\mids(\{a_i,b_i\})$, and the latter
has size at most $k$ as $\w(T_i,\mu_i) \leq k$. For all other edges $e$ of
$T_i$, $i=1,2$, $\mids(e)$ is exactly the same in $T$ and in $T_i$, since
if $e \in E(T_i^v)$ then $v \in \mids(e)$ in both $T$ and $T_i$, and if $e
\in E(T_i \setminus T_i^v)$ then $v \notin \mids(e)$ in both $T$ and
$T_i$.

If $G_1$ and $G_2$ share no vertices, we can merge two branch
decompositions $(T_1,\mu_1)$ and $(T_2,\mu_2)$ by subdividing a pair of
arbitrary edges, without increasing the width. \findem

\begin{lemma}[Fomin and Thilikos~\cite{FoTh06}]
\label{lem:sum2} Let $G_1$ and $G_2$ be graphs with one edge $f$ in
common. Then $\bw(G_1 \cup G_2) \leq \max \{\bw(G_1),\bw(G_2),2\}$.
Moreover, if both endvertices of $f$ have degree at least two in at least
one of the graphs, then $\bw(G_1 \cup G_2) = \max \{\bw(G_1),\bw(G_2)\}$.
\end{lemma}

\begin{figure}[t]
\begin{center}
\includegraphics[width=12.0cm]{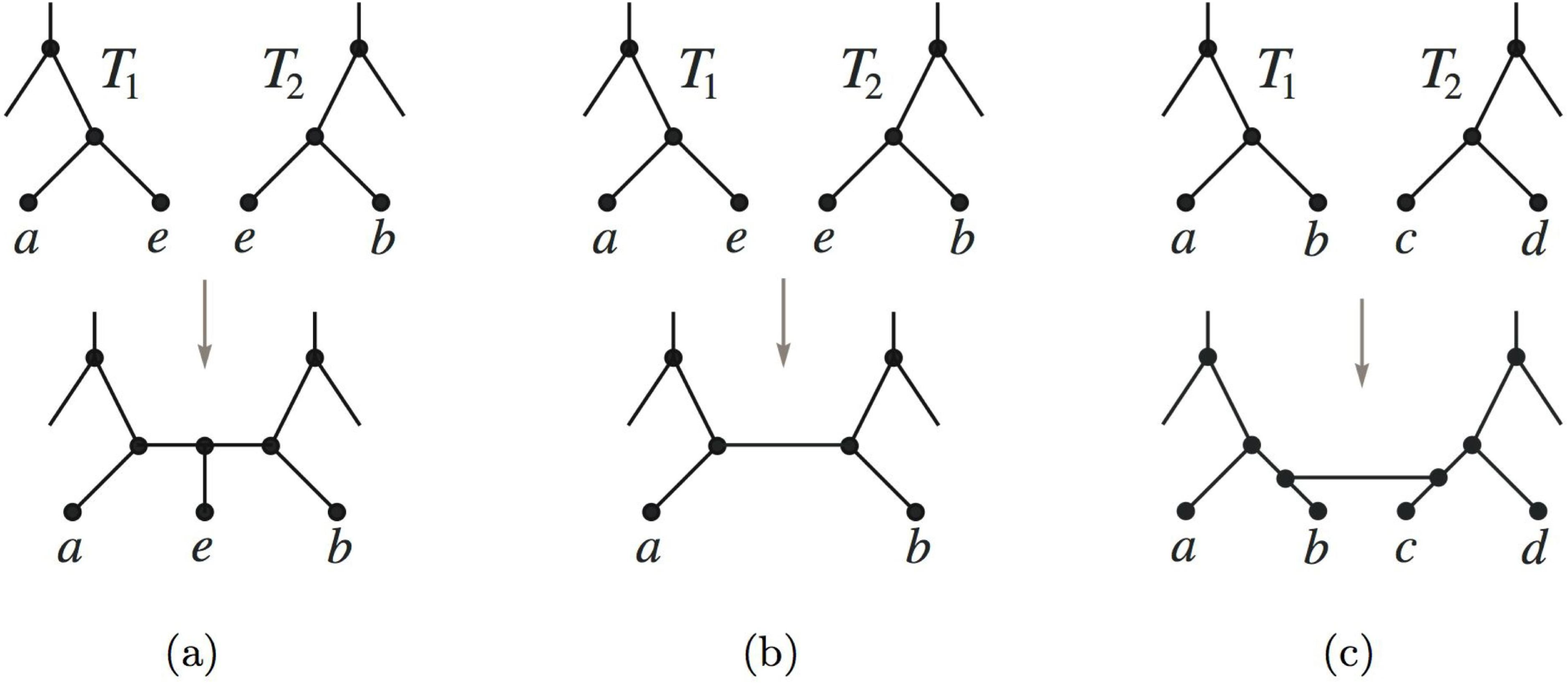}
\end{center}

\vspace{-0.7cm} \caption{Merging branch decompositions $(T_1,\mu_1)$ and
$(T_2,\mu_2)$ of two components $H_1$ and $H_2$ in a polyhedral
decomposition $(\mathcal{G},A)$ of $G=(V,E)$. There are three cases: (a)
$H_1$ and $H_2$ share two vertices $v_1$, $v_2$ and the edge
$e=\{v_1,v_2\}$ is in $E$; (b) $H_1$ and $H_2$ share two vertices $v_1$,
$v_2$ and $e=\{v_1,v_2\}$ is \emph{not} in $E$; (c) $H_1$ and $H_2$ share
one vertex $v$.} \label{fig:tree1}
\end{figure}

It remains only to show how to merge the branch decompositions
$(T_1,\mu_1)$, $(T_2,\mu_2)$ of two graphs $H_1$, $H_2$ in $\mathcal{G}$.
We distinguish four cases:\vspace{0.2cm}
\begin{itemize}
\item[(a)] $H_1$ and $H_2$ share two
vertices $v_1$, $v_2$, and the edge $e=\{v_1,v_2\} \in E(G)$. We take the
leaves in $T_1$ and $T_2$ corresponding to $e$, we identify them, and we
add a new edge whose leave corresponds to $e$ (see
Figure~\ref{fig:tree1}(a)). \vspace{-0.12cm}
\item[(b)] $H_1$ and $H_2$ share two
vertices $v_1$, $v_2$, and the edge $e=\{v_1,v_2\} \notin E(G)$. We take
the leaves in $T_1$ and $T_2$ corresponding to $e$, we identify them, and
we dissolve the common vertex (see Figure~\ref{fig:tree1}(b)).
\vspace{-0.12cm}
\item[(c)] $H_1$ and $H_2$ share one vertex $v$. We take two edges $b,c$ in
$T_1,T_2$ whose leaves correspond to edges containing $v$, we subdivide
them and add a new edge between the newly created vertices (see
Figure~\ref{fig:tree1}(c)). \vspace{-0.12cm}
\item[(d)] $H_1$ and $H_2$ share no vertices. We do the construction of case~(c) for any two edges of the two branch decompositions.
\end{itemize}
\vspace{0.2cm} The above construction does not increase the branchwidth by
Lemmata~\ref{lem:sum1} and~\ref{lem:sum2}.

\findem

\begin{observation} \label{obs:addVertices} Let $G=(V,E)$ be
a graph, and let $A \subseteq V$. Given a branch decomposition $(T',\mu')$
of $G[V \setminus A]$, we can obtain a branch decomposition $(T,\mu)$ of
$G$ with $\w(T,\mu) \leq \w(T',\mu') + |A|$ recursively as follows: First,
for each edge $\{u,v\} \in E(G)$ with $u \in V \setminus A$ and $v \in A$,
we choose an edge $e \in G[V \setminus A]$ containing $u$, and we replace
the leaf of $T'$ corresponding to $e$ with two incident pendant edges
whose leaves correspond to edges $\{u,v\}$ and $e$, respectively. Finally,
for each edge $\{u,v\} \in E(G)$ with $u,v \in A$, we take and arbitrary
edge of $T'$, subdivide it, and add a new edge whose leave corresponds to
edge $\{u,v\}$. It can be easily checked that the size of the middle sets
has increased by at most $|A|$.
\end{observation}



Given an embedded graph $G$ and a carving decomposition $(T,\mu)$ of its
medial graph $M_G$, we define a \emph{radial decomposition} $(T^*,\mu^*)$
of the dual graph $R_G$,  where $T^*=T$ and $\mu^*$ is a bijection from
the leaves of $T$ to the set of faces of $R_G$ defined as follows: for
each edge $e \in E(T)$, $\mu^*(e)=f$, where $f$ is the face in $R_G$
corresponding to the vertex $u_f \in V(M_G)$ such that $\mu(e)=u_f$. Each
edge $e \in E(T^*)$ partitions the faces of $R_G$ into two sets $F_1$ and
$F_2$. We define the \emph{border set} of $e$, denoted $\bor(e)$, as the
set of edges of $R_G$ that belong to both $F_1$ and $F_2$. Note that $F_1$
and $F_2$ may intersect also in vertices, not only in edges.

If $(T,\mu)$ is a bond carving decomposition of $M_G$, then the associated
radial decomposition (also called \emph{bond}) has nice connectivity
properties. Indeed, in a bond carving decomposition, every cut set
partitions the vertices of $M_G$ into two subsets $V_1,V_2$ such that both
$M_G[V_1]$ and $M_G[V_2]$ are non-null and connected. This property, seen
in the radial decomposition of $R_G$, implies that each edge $e \in
E(T^*)$ corresponds to a partition of the faces of $R_G$ into two sets
$F_1$ and $F_2$, namely \emph{black} and \emph{white} faces (naturally
partitioning the edges into \emph{black}, \emph{white}, and \emph{grey}),
such that it is possible to reach any black (resp. white) face from any
black (resp. white) face by only crossing black (resp. white) edges. In
other words, the union of all black (resp. white) faces and edges is a
connected set. 

\begin{observation}
\label{obs:natural}Recall that all the faces of a radial graph $R_G$ are
\emph{tiles}, that is, each face has exactly 4 edges. Also, each one of
those tiles corresponds to a pair of dual edges $e$ and $e^*$ of $G$ and
$G^*$, respectively. Given a carving decomposition $(T,\mu)$ of $M_G$ (or
equivalently, a radial decomposition $(T^*,\mu^*)$ of $R_G$), one can
obtain in a natural way branch decompositions of $G$ and $G^*$ by
redefining the bijection $\mu$ from the leaves of $T$ to the edges of $G$
(or $G^*$) that correspond to the faces of $R_G$.
\end{observation}

\section{Some topological results}
\label{sec:topological_lemmas} In this section we state two topological
lemmata and some definitions that will be used in
Section~\ref{sec:sphere_cut}. Given a collection ${\cal S}$ of sets, we
denote their union by $\cupall{\cal S}=\bigcup_{S\in{\cal S}}S$.


Given a graph $G$ embedded in a surface of Euler genus $\gamma$, its dual
$G^*$ and a spanning tree $C^*$ of $G^*$, we call $C = \{e\in E(G)\ |\ e^*
\in E(C^*)\}$ a \emph{spanning cotree} of $G$. We define a
\emph{tree-cotree partition} (cf.~\cite{Epp03}) of an embedded graph $G$
to be a triple $(T,C,X)$ where $T$ is a spanning tree of $G$, $C$ is a
spanning cotree of $G$, $X\subseteq E(G)$, and the three sets $E(T)$, $C$,
and $X$ form a partition of $E(G)$. Eppstein proved~\cite[Lemma
3.1]{Epp03} that if $T$ and $C^*$ are forests such that $E(T)$ and $C$ are
disjoint, we can make $T$ become part of a spanning tree $T'$ and $C$
become part of a spanning cotree disjoint from $T'$, extending $T$ and $C$
to a tree-cotree decomposition. We can now announce the following lemma
from~\cite[Lemma 3.2]{Epp03}.

\begin{lemma}[Eppstein~\cite{Epp03}]
\label{lem:topo1} If $(T,C,X)$ is a tree-cotree decomposition of a graph
$G$ embedded in a surface of Euler genus $\gamma$, then $|X|=O(\gamma)$.
\end{lemma}



Let $\Sigma$ be a surface and let $\mathcal{N}$ be a finite collection of
$O$-arcs in $\Sigma$ pairwise intersecting at a finite zero-dimensional
subsets (i.e., points) of $\Sigma$. For a point $p \in \Sigma$, let
$\mathcal{N}(p)$ be the number of $O$-arcs in $\mathcal{N}$ containing
$p$, and let $P(\mathcal{N}) = \{p \in \Sigma : \mathcal{N}(p) \geq 2\}$;
 note that by assumption $P(\mathcal{N})$ is a finite set of
 points of $\Sigma$. Then we define
$$
\theta(\mathcal{N}) = \sum_{p \in P(\mathcal{N})} (\mathcal{N}(p)-1)\ .
$$

\begin{lemma}\label{lemma: O(g)}
Let $\Sigma$ be a surface without boundary with $\gamma(\Sigma)= \gamma$.
Let $\mathcal{N}$ be a collection of $O(\gamma)$ $O$-arcs in $\Sigma$
pairwise intersecting at finite zero-dimensional subsets of $\Sigma$, and
such that $\Sigma \setminus \cupall\mathcal{N}$ has two connected
components. Then $\theta(\mathcal{N}) = O(\gamma).$


\end{lemma}
\dem In order to prove the lemma, we define from $\mathcal{N}$ the
following (multi)graph $H$ embedded in $\Sigma$: we first add a vertex
$v_p$ in $H$ for every point $p$ in $\Sigma$ such that $\mathcal{N}(p)
\geq 2$. We call such points \emph{repeated}. We now distinguish four
cases according to the number of repeated points in an $O$-arc. First, for
each $O$-arc $N$ with at least three repeated points, we order cyclically
the repeated points in $N$, and the same ordering applies to the
corresponding vertices in $H$. Then, we add an edge in $H$ between each
two consecutive vertices in that ordering. For each $O$-arc with exactly
two repeated points $p$ and $q$, we add two parallel edges in $H$ between
$v_p$ and $v_q$. For each $O$-arc with exactly one repeated point $p$, we
add in $H$ a loop at vertex $v_p$. Finally, for each $O$-arc $N$ with no
repeated points, we add to $H$ a new vertex $v_N$ with a loop. Visually,
$H$ is the graph embedded in $\Sigma$ corresponding to the union of the
$O$-arcs in $\mathcal{N}$.

In order to prove the result, by the construction of $H$ it is enough to
prove that $\sum_{v\in V(H)}(\dd_H(v)-2)= O(\gamma)$.
By assumption, $H$ separates $\Sigma$ into two connected components
$\Sigma'$ and $\Sigma''$. Let $H_1, H_2,\dots, H_r$ be the maximal
connected subgraphs of $H$. In particular, $r \leq |\mathcal{N}| =
O(\gamma)$ by hypothesis. Some of these connected subgraphs may be
incident with $\Sigma'$ but not with $\Sigma''$, or vice-versa.
Additionally, there is at least one connected subgraph $H_i$ incident with
both connected components. Without loss of generality we assume that the
subgraphs $H_1,H_2,\dots, H_p$ are incident only with $\Sigma'$,
$H_{p+1},\dots,H_q$ are incident with both components, and
$H_{q+1},\dots,H_{r}$ are incident only with $\Sigma''$. It is clear that
there exists a path joining a vertex in $H_{i}$ with a vertex in $H_{i+1}$
if $1\leq i \leq q-1$ or $p+1\leq i \leq r-1$.

From graphs $H_1,H_2,\dots, H_p,\dots,H_q$ (the ones which are incident
with $\Sigma'$) we construct a new graph $G_1$ in the following inductive
way: we start taking $H_{q}$ and $H_{q-1}$, and a path joining a vertex in
$H_{q}$ to a vertex in $H_{q-1}$. This path exists because $H_{q}$ and
$H_{q-1}$ are incident with $\Sigma'$. Consider the graph obtained from
$H_{q}$ and $H_{q-1}$ by adding an edge that joins this pair of vertices.
Then, we delete $H_{q}$ and $H_{q-1}$ from the initial list and add this
new connected graph. This procedure is done $q-1$ times. At the end, we
obtain a connected graph $G'$ incident with both $\Sigma'$ and $\Sigma''$
where each vertex has degree at least three. Finally, we apply the same
procedure with $G',H_{q+1},\dots,H_{r}$, obtaining a connected graph $G$.
Observe also that
$$\sum_{v\in V(H)}(\dd_H(v)-2)\ \leq\ \sum_{v\in V(G)}(\dd_G(v)-2)\ <\ \sum_{v\in V(G)}\dd_G(v)\ =\ 2|E(G)|\ .$$
In what follows, we obtain upper bounds for $2|E(G)|$. Observe that $H$
defines a pair of faces over $\Sigma$, not necessarily disks. In the
previous construction of $G$, every time we add an edge we either
subdivide a face into two parts or not. Consequently, the number of faces
that $G$ defines over $\Sigma$ is at most $2+|\mathcal{N}|$. The next step
consists in reducing the surface in the following way: let $f$ be a face
determined by $G$ over $\Sigma$. If $f$ is contractible, we do nothing. If
it is not, there is a non-contractible cycle $\mathbb{S}^1$ contained on
$f$. Let $\Sigma_1$ be the connected component of
$\Sigma$\Leftscissors$\mathbb{S}^1$ which contains $G$. Then $G$ defines a
decomposition of $\Sigma_1$, $\gamma(\Sigma_1)\leq \gamma$, and the number
of faces has been increased by at most one. Observe that for each
operation \Leftscissors$\,$ we reduce the Euler genus and we create at
most one face. As the Euler genus is finite, so is the number of
\Leftscissors$\,$ operations. This gives rise to a surface $\Sigma_s$ with
$\gamma(\Sigma_s)\leq\gamma$, and such that all faces determined by $G$
are contractible. Additionally, the number of faces that $G$ determines
over $\Sigma_s$ is smaller than $2+|\mathcal{N}|+\gamma$.

$G$ defines a map on $\Sigma_s$ (i.e., all faces are contractible), and
consequently we can apply Euler's formula. Then
$|F(G)|+|V(G)|=|E(G)|+2-\gamma(\Sigma_s)$. Then, as $|F(G)|\leq
2+|\mathcal{N}|+\gamma$, we obtain that
$|E(G)|+2-\gamma(\Sigma_s)=|V(G)|+|F(G)|\leq
|V(G)|+2+|\mathcal{N}|+\gamma$. The degree of each vertex is at least
three, and thus $3|V(G)|\leq 2|E(G)|$. Substituting this condition in the
previous equation, we obtain
$$|E(G)|+2-\gamma(\Sigma_s)\ \leq\ |V(G)|+2+|\mathcal{N}|+\gamma\ \leq\ \frac{2}{3}|E(G)|+2+|\mathcal{N}|+\gamma.$$
Isolating $|E(G)|$, we get that $2|E(G)|\leq
6|\mathcal{N}|+6\gamma(\Sigma_s)+6\gamma\leq 6|\mathcal{N}|+12\gamma$. As
by hypothesis $|\mathcal{N}|=O(\gamma)$, the previous bound yields the
desired result.\findem

%


\section{Surface cut decompositions} \label{sec:sphere_cut}
Sphere cut decompositions have been introduced as a combinatorial concept in~\cite{SeymourT94} and were used for the first time in~\cite{DornPBF10effi}
to analyze the running time of algorithms based on dynamic programming
over branch decompositions on planar
graphs (see also~\cite{DFT07,SaTh10,DFT08}. In this section we generalize
sphere cut decompositions to graphs on surfaces; we call them
\emph{surface cut decompositions}.

\begin{definition} \label{def:surface_cut} Given a graph $G$
embedded in a surface $\Sigma$ with $\gamma(\Sigma)=\gamma$, a
\emph{surface cut decomposition} of $G$ is a branch decomposition
$(T,\mu)$ of $G$ such that there exists a polyhedral decomposition
$(\mathcal{G},A)$ of $G$ with the following property: for each edge $e \in
E(T)$, either $|\mids(e) \setminus A| \leq 2$, or there exists a graph $H
\in \mathcal{G}$ such that\vspace{-0.0000cm}
\begin{itemize}
\item $\mids(e) \setminus A \subseteq V(H)$;\vspace{-0.12cm}
\item the vertices in $\mids(e) \setminus A$ are contained
in a set $\mathcal{N}$ of $O(\gamma)$ nooses of $\Sigma$ pairwise
intersecting only at subsets of $\mids(e) \setminus A$.\vspace{-0.12cm}
\item $\theta(\mathcal{N})= O(\gamma)$.
\vspace{-0.12cm}
\item  $\Sigma \setminus \cupall \mathcal{N}$ contains
exactly two connected components, such that the graph $G_{e} \setminus A$
is embedded in the closure of one of them.
\end{itemize}
\end{definition}

Note that a sphere cut decomposition is a particular case of a surface cut
decomposition when $\gamma=0$, by taking $A = \emptyset$, $\mathcal{G}$
containing only the graph $G$ itself, and all the vertices of each middle
set contained in a single noose. We provide now an algorithm to construct
a surface graph decomposition of an embedded graph. The
proof of Theorem~\ref{teo:surface_cut} 
uses Proposition~\ref{prop:poly} and all the results of Sections~\ref{sec:width} and~\ref{sec:topological_lemmas}.\\
\vspace{0.2cm}

\begin{algorithm}[htb]
{\small \caption{Construction of a surface cut decomposition of an
embedded graph $G$} \label{alg:surface}
\begin{algorithmic} 
\REQUIRE{An embedded graph $G$.}

\ENSURE{A surface cut decomposition of $G$.}
\medskip

\STATE{Compute a polyhedral decomposition $(\mathcal{G},A)$ of $G$, using
Algorithm~\ref{alg:polyhedral}.}

\FOR{each component $H$ of $\mathcal{G}$}
    \STATE{1. Compute a branch decomposition $(T'_H,\mu'_H)$ of $H$, using~\cite[Theorem 3.8]{Ami01}.}
    \STATE{2. Transform $(T'_H,\mu'_H)$ to a carving decomposition $(T^c_H,\mu^c_H)$ of the medial graph $M_{H}$, using Lemma~\ref{lem:carving12}.}
    \STATE{3. Transform $(T^c_H,\mu^c_H)$ to a \emph{bond} carving decomposition $(T^b_H,\mu^b_H)$ of $M_{H}$, using~\cite{SeymourT94}.}
    \STATE{4. Transform $(T^b_H,\mu^b_H)$ to a branch decomposition $(T_H,\mu_H)$ of $H$, using Observation~\ref{obs:natural}.}
\ENDFOR

\STATE{Construct a branch decomposition $(T,\mu)$ of $G$ by merging, using
Lemma~\ref{lem:bw2}, the branch decompositions $\{(T_H,\mu_H)\ |\ H \in
\mathcal{G} \}$, and by adding the edges of $G$ with at least one
endvertex in $A$, using Observation~\ref{obs:addVertices}.}

\RETURN{$(T,\mu)$.}
\end{algorithmic}
}
\end{algorithm}

\vspace{-0.4cm}

\begin{theorem}
\label{teo:surface_cut} Given a graph $G$ on $n$ vertices embedded in a
surface of Euler genus $\gamma$, with $\bw(G)\leq k$,
Algorithm~\ref{alg:surface} constructs, in
 $2^{3k+O(\log k)} \cdot n^3$ steps,
a surface cut decomposition $(T,\mu)$ of $G$ of width at most $27k +
O(\gamma)$.
\end{theorem}
\dem We prove, in this order, that the output $(T,\mu)$ of
Algorithm~\ref{alg:surface} is indeed a surface cut decomposition of $G$,
then that the width of $(T,\mu)$ is at most $27 \bw(G) + O(\gamma)$,  and
finally the claimed running time.

\paragraph{$(T,\mu)$ is a surface cut decomposition of $G$.} We shall prove that all the properties of
Definition~\ref{def:surface_cut} are fulfilled. First note that, as
$(\mathcal{G},A)$ is a polyhedral decomposition of $G$, we have that
$|A|=O(\gamma)$.

By construction, it is clear that $(T,\mu)$ is a branch decomposition of
$G$. In $(T,\mu)$, there are some edges that have been added in the last
step of Algorithm~\ref{alg:surface}, in order to merge branch
decompositions of the graphs in $\mathcal{G}$, with the help of
Lemma~\ref{lem:bw2}. Let $e$ be such an edge. Since $(\mathcal{G},A)$ is a
polyhedral decomposition of $G$, any two graphs in $\mathcal{G}$ share at
most two vertices, hence $|\mids(e)\setminus A| \leq 2$.

All other edges  of $(T,\mu)$ correspond to an edge of a branch
decomposition of some polyhedral component $H \in \mathcal{G}$. Let
henceforth $e$ be such an edge. Therefore, $\mids(e) \setminus A \subseteq
V(H)$. To complete this part of the proof, we prove in a sequence of three
claims that the remaining conditions of Definition~\ref{def:surface_cut}
hold.

\begin{claim}
\label{claim:1} The vertices in $\mids(e) \setminus A$ are contained in a
set $\mathcal{N}$ of $O(\gamma)$ nooses.
\end{claim}
\dem The proof uses the tree-cotree partition defined in
Section~\ref{sec:topological_lemmas}.

Recall that $e$ is an edge that corresponds to a branch decomposition
$(T_H,\mu_H)$ of a polyhedral component $H$ of $\mathcal{G}$. The branch
decomposition $(T_H,\mu_H)$ of $H$ has been built by
Algorithm~\ref{alg:surface} from a bond carving decomposition of its
medial graph $M_H$, or equivalently from a bond radial decomposition of
its radial graph $R_H$. Due to the fact that the carving decomposition of
$M_H$ is bond, edge $e$ partitions the vertices of $M_H$ into two sets --
namely, \emph{black} and \emph{white} vertices -- each one inducing a
connected subgraph of $M_H$. There are three types of edges in $R_H$:
\emph{black}, \emph{white}, and \emph{grey}, according to whether they
belong to faces of the same color (black or white) or not. Therefore, the
corresponding black and white faces also induce connected subgraphs of
$R_H$, in the sense that it is always possible to reach any black (resp.
white) face from any black (resp. white) face only crossing black (resp.
white) {\sl edges}.

Let $F$ be the set of grey edges of $R_H$. Since each edge of $R_H$
contains a vertex from $H$ and another from $H^*$, the vertices in
$\mids(e)$ are contained in $R_H[F]$, so it suffices to prove that
$R_H[F]$ can be partitioned into a set of $O(\gamma)$ cycles (possibly
sharing some vertices). Note that each cycle in the radial graph $R_H$
corresponds to a noose in the surface.

To this end, first note that in $R_H[F]$ all vertices have even degree.
Indeed, let $v \in V(R_H[F])$, and consider a clockwise orientation of the
edges incident with $v$ in $R_H[F]$. Each such edge alternates from a
black to a white face, or viceversa, so beginning from an arbitrary edge
and visiting all others edges in the clockwise order, we deduce that the
number of edges incident with $v$ is necessarily even.

Therefore, $R_H[F]$ can be partitioned into a set of cycles. Let us now
bound the number of such cycles. Since the subgraph induced by the black
(resp. white) faces of $R_H$ is connected, we can consider in $M_H$ a
spanning tree $T^{*}_B$ (resp. $T_W^*$) corresponding to the black (resp.
white) faces of $R_H$. Merge both trees by adding a new edge $e_0^*$, and
let $T^*$ be the resulting tree. Let $T$ be a spanning tree of $R_H$
disjoint from $T^*$ (in the sense that there is no pair of dual edges $e$
and $e^*$ with $e \in E(T)$ and $e^* \in E(T^*)$); such a spanning tree
exists by~\cite[Lemma 3.1]{Epp03}. Now consider the tree-cotree partition
$(T,T^*,X)$, where $X$ is the set of edges of $R_H$ that are neither in
$T$ nor in $T^*$.

Each edge of $T^*$, except $e_0^*$, corresponds to two faces of $R_H$ of
the same color. Therefore, the set $F \in E(R_H)$ of edges separating
faces of different color is contained in $T\cup\{e_0\}\cup X$. Since $T$
is a tree, each cycle of $R_H[F]$ uses at least one edge in $\{e_0\}\cup
X$. Therefore, $R_H[F]$ can be partitioned into at most $1 + |X|$ cycles.
The result follows from the fact that $(T,T^*,X)$ is a tree-cotree
partition, and therefore $|X|=O(\gamma)$ by Lemma~\ref{lem:topo1}. \findem

\begin{claim}
\label{claim:2} Let $\mathcal{N}$ be the set of nooses constructed in the
proof of Claim~\ref{claim:1}. Then $\cupall \mathcal{N}$ separates
$\Sigma$ into two connected components.
\end{claim}
\dem By Claim~\ref{claim:1}, the vertices in $\mids(e) \setminus A$ are
contained in $\cupall \mathcal{N}$. The claim holds from the fact that for
each component $H$ of $\mathcal{G}$, $(T^b_H,\mu^b_H)$ is a bond carving
decomposition of $M_{H}$, and by taking into account the
discussion before Observation~\ref{obs:natural}.\findem  

Note that the collection of nooses constructed in the proof of
Claim~\ref{claim:1} is finite and its elements pairwise intersect only at
subsets of $\mids(e) \setminus A$, as required. In particular, for this
collection $\mathcal{N}$ of nooses, the parameter $\theta(\mathcal{N})$ is
well-defined (see Section~\ref{sec:topological_lemmas}).

\begin{claim}
\label{claim:3} Let $\mathcal{N}$ be the set of nooses constructed in the
proof of Claim~\ref{claim:1}. Then $\theta(\mathcal{N}) = O(\gamma).$
\end{claim}
\dem By Claim~\ref{claim:2}, $\cupall \mathcal{N}$ separates
$\Sigma$ into two connected components. 
The claim then holds by Lemma~\ref{lemma: O(g)}. \findem 

\paragraph{The width of $(T,\mu)$ is at most $27 \cdot \bw(G) +
O(\gamma)$.} For simplicity, let $k=\bw(G)$. By
Proposition~\ref{prop:poly}, each polyhedral component $H$ is a minor of
$G$, hence $\bw(H)\leq k$ for all $H \in \mathcal{G}$. In Step 1 of
Algorithm~\ref{alg:surface}, we compute a branch decomposition
$(T'_H,\mu'_H)$ of $H$ of width at most $k'= \frac{9}{2}k$, using Amir's
algorithm~\cite[Theorem 3.8]{Ami01}. In Step 2, we transform
$(T'_H,\mu'_H)$ to a carving decomposition $(T^c_H,\mu^c_H)$ of the medial
graph $M_{H}$ of $H$ of width at most $12k'$, using
Lemma~\ref{lem:carving12}. In Step 3, we transform $(T^c_H,\mu^c_H)$ to a
{\sl bond} carving decomposition $(T^b_H,\mu^b_H)$ of $M_{H}$ of width at
most $12k'$, using the algorithm of~\cite{SeymourT94}. Then, using
Observation~\ref{obs:natural}, we transform in Step 4 $(T^b_H,\mu^b_H)$ to
a branch decomposition $(T_H,\mu_H)$ of $H$. By the proof of
Claim~\ref{claim:1}, the discrepancy between $\w(T_H,\mu_H)$ and
$\w(T^b_H,\mu^b_H)/2$ is at most the bound provided by Lemma~\ref{lemma:
O(g)}, i.e., $O(\gamma)$. Therefore, $\w(T_H,\mu_H) \leq 6k'+O(\gamma)=27k
+ O(\gamma)$, for all $H \in \mathcal{G}$.

Then, we merge the branch decompositions of the polyhedral components,
using Lemma~\ref{lem:bw2}, and finally we add the edges of $G$ with at
least one endvertex in $A$, using Observation~\ref{obs:addVertices}, to
obtain a branch decomposition $(T,\mu)$ of $G$.


Combining the discussion above with Lemmata~\ref{lem:bw1}
and~\ref{lem:bw2} and Observation~\ref{obs:addVertices}, and using that
$|A|= O(\gamma)$, we get that
\begin{eqnarray*} \w(T,\mu)&\leq &\max \{ 2, \{ \w(T_H,\mu_H)\ |\ H \in
\mathcal{G}\} \} + |A|\\
& \leq & 27k + O(\gamma) + |A|\\
& = & 27k + O(\gamma).
\end{eqnarray*}

\paragraph{Algorithm~\ref{alg:surface} runs in $2^{3k+O(\log
k)}\cdot n^3$ time.}

We analyze sequentially the running time of each step. First, we compute a
polyhedral decomposition of $G$ using Algorithm~\ref{alg:polyhedral} in
$O(n^3)$ steps, by Proposition~\ref{prop:poly}. Then, we run Amir's
algorithm in each component in Step 1, which takes $O(2^{3k}k^{3/2}n^2)$
time~\cite[Theorem 3.8]{Ami01}. We would like to stress that this step is
the only non-polynomial procedure in the construction of surface cut
decompositions. Step 2 can be done in linear time by
Lemma~\ref{lem:carving12}. Step 3 can be done in $O(n^2)$
time~\cite{SeymourT94}. Step 4 takes linear time by
Observation~\ref{obs:natural}. Merging the branch decompositions can
clearly be done in linear time. Finally, since any two elements in
$\mathcal{G}$ share at most two vertices, the overall running time is the
claimed one.
\findem

\section{Upper-bounding the size of the tables}
\label{sec:upperbounds}


In this section we show that by using surface cut decompositions in order
to solve connected packing-encodable problems in surface-embedded graphs,
one can guarantee single-exponential upper bounds on the size of the
tables of dynamic programming algorithms. Then Theorem~\ref{teo:finals}
follows directly by the definition of a connected packing-encodable
problem and the following lemma.

%

\begin{lemma}
\label{lem:jbf} Let $G$ be a graph embedded in a surface $\Sigma$ without
boundary and Euler genus $\gamma$, and let $(T,\mu)$ be a surface cut
decomposition of $G$ of width at most $k$. Then for every $e\in E(T)$,
$|\Psi_{G_{e}}(\mids(e))|= \gamma^{O(\gamma)}\cdot k^{O(\gamma)}\cdot
\gamma^{O(k)}$.
\end{lemma}

Before we give the proof of the above lemma, we first  need to define
formally the notion of  non-crossing partitions on surfaces with boundary
and then to prove some lemmata that
combine elements from topology and combinatorics.\\

\begin{figure}[h!tb]
\begin{center}
\includegraphics[width=11.0cm]{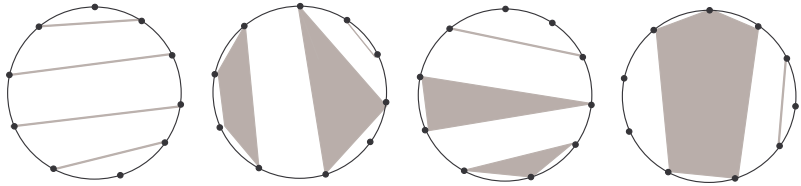}
\caption{Non-crossing partitions on a disk, which enumerate the number of
partial solutions on planar graphs when using sphere cut
decompositions.}\label{fig:catalan}
\end{center}
\end{figure}

 A \emph{non-crossing partition}
of a set of size $k$, from a combinatorial point of view, is a partition
of the set $\{1,2,\dots ,k\}$ with the following property: if $\{a,b,c,d\}
\subseteq \{1, 2,\dots ,k\}$ with $1\leq a<b<c<d\leq k$ and some subset in
the partition contains $a$ and $c$, then no other subset contains both $b$
and $d$. One can represent such a partition on a disk by placing $k$
points on the boundary of the disk, labeled consecutively, and drawing
each subset as a convex polygon (also called \emph{block}) on the points
belonging to the subset. Then, the ``non-crossing'' condition is
equivalent to the fact that the blocks are pairwise disjoint. See
Figure~\ref{fig:catalan} for some examples.

The enumeration of non-crossing partitions on a disk is one of the first
non-trivial problems in enumerative combinatorics: it is well-known (see
e.g.~\cite{FlajoletSedgewig:analytic-combinatorics}) that the the number
of non-crossing partitions of $\{1,2,\dots,k\}$ on a disk is equal to the
Catalan number $C(k)=\frac{1}{k+1}\binom{2k}{k}
\sim\frac{4^{k}}{k^{3/2}\sqrt{\pi}}= O(4^{k})$. This is the main
combinatorial property exploited to obtain single-exponential dynamic
programming algorithms on planar graphs using sphere cut
decompositions~\cite{DornPBF10effi,SeymourT94,SaTh10}.



The generalization of the notion of non-crossing partition to surfaces
of higher genus is not as straignforward as in the case of the disk, and must be defined carefully. 
We consider pairs $(\Sigma,S)$
where $\Sigma$ is a surface whose boundary has $\beta\left(\Sigma\right)$
connected components, each one homeomorphic to a simple circle, and  $S$
is a set of vertices on this boundary. A {\em partition family} for the
pair $(\Sigma,S)$ is a collection $\mathfrak{B}$ of mutually
non-intersecting connected subsets of $\Sigma$,
 such that each vertex in $S$
belongs to some set in $\mathfrak{B}$.

Actually the concept of a partition family is not enough for our purposes,
 as we have to incorporate the presence of
the set of vertices $A$ arising from a polyhedral decomposition. This set
of vertices plays is some sense the role of {\sl apices} in Graph Minors
theory, and this is why we also call these vertices \emph{apices}. For
this reason we consider pairs of the form $(\Sigma\cup \Gamma_{A},S\cup
A)$ where $\Sigma$ is a surface with boundary, $S$ is a set of vertices on
this boundary, $A$ is a vertex set not on the boundary (the apices), and
$\Gamma_{A}$ is the closed set containing the points of the graph $C_{A}$
obtained if we take a complete graph with $A$ as vertex set and add to it
$S$ together with all edges between the vertices of $A$ and $S$. We
require that $\Gamma_{A}\cap \Sigma=S$ and we see the set $\Gamma_{A}$ as
``flying above'' the surface $\Sigma$. That way, we call the edges of
$C_{A}$ {\em flying edges}, and we treat them as {\sl closed subsets} of
$\Gamma_{A}$ by adding to them the two endpoints of their boundary. We
use the notation $\Sigma^{A}$ to denote $\Sigma\cup\Gamma_{A}$ (clearly
$\Sigma=\Sigma^{\emptyset}$).
To extend the definition of partition family, we take a partition family
$\mathfrak{B}_{\Sigma}$ of $\Sigma$ and, on top of it, we consider a set
$\mathfrak{E}_{A}$ of flying edges where each apex is incident with some
edge in $\mathfrak{E}_{A}$. An {\em extended partition family} for
$(\Sigma^{A},S)$  is a collection $\mathfrak{B}_{\Sigma^{A}}$ of subsets
of $\Sigma^{A}\cup S$ defined as
$$\mathfrak{B}_{\Sigma^{A}}=\{C\mid C\mbox{~is a connected component of the set $\cupall (%
\mathfrak{B}_{\Sigma}\cup
\mathfrak{E}_{A}%
)$}%
\}\ ,$$
%
where $\mathfrak{B}_{\Sigma}$ and $\mathfrak{E}_{A}$ are taken as before.
For simplicity, we may drop the index of a collection
$\mathfrak{B}_{\Sigma}$ or
 $\mathfrak{B}_{\Sigma^{A}}$
when it is clear from the context whether it refers to $\Sigma$ or to
$\Sigma^{A}$.

Notice that each partition family $\mathfrak{B}$ for $(\Sigma^{A},S\cup
A)$ defines a partition of $S\cup A$ as follows.
$${\cal R}(\mathfrak{B})=\{(S\cup A)\cap B\mid B\in \mathfrak{B}  \}.$$

We say that two extended partition families $\mathfrak{B}_{1}$ and
$\mathfrak{B}_{2}$ for $(\Sigma^{A},S\cup A)$ are {\em equivalent} if
${\cal R}(\mathfrak{B}_{1})={\cal R}(\mathfrak{B}_{2})$ and we denote it
by $\mathfrak{B}_{1}\equiv \mathfrak{B}_{2}$. The set of the {\em
non-crossing partitions with apices} of the set $S\cup A$ (where $S$ and
$A$ are vertices embedded in $\Sigma^{A}$ as before), denoted by
$\Pi_{\Sigma^{A}}(S\cup A)$,  is the set of equivalence classes of the
extended partition families for $(\Sigma^{A},S\cup A)$ with respect to the
relation $\equiv$.  
%
%

We define $\Pi_{\Sigma}(S)=\Pi_{\Sigma^{\emptyset}}(S\cup \emptyset)$, and
note that, if $\Sigma$ is the disk and $|S|=k$, then $|\Pi_{\Sigma}(S)|$
is the $k$-th Catalan number and therefore $|\Pi_{\Sigma}(S)|=O(4^{k})$.
The asymptotic enumeration of $|\Pi_{\Sigma}(S)|$ for general surfaces is
quite a complicated problem. However, its behavior for surfaces $\Sigma$
where $\gamma(\Sigma)$ and $\beta(\Sigma)$ are bounded is not
significantly different from the disk in what concerns its exponential
growth. In particular it holds that $\lim_{|S|\rightarrow \infty}
|\Pi_{\Sigma}(S)|^{1/|S|}=4$ and this is a consequence of the following
enumerative result from~\cite{RST10_comb_Arxiv}.

\begin{theorem}
\label{theorem: non-crossing-partition-final} Let $\Sigma$ be a surface
with boundary. Then the number $|\Pi_{\Sigma}(S)|$, for $|S|=k$, verifies
\begin{equation}\label{eq:thm-111}
|\Pi_{\Sigma}(S)|\leq _{k\rightarrow
\infty}\frac{C(\Sigma)}{\Gamma\left(3/2\gamma(\Sigma)+\beta(\Sigma)-3\right)}\cdot
k^{3/2\gamma(\Sigma)+\beta(\Sigma)-4} \cdot 4^k\ ,
\end{equation}
where $C(\Sigma)$ is a function depending only on $\Sigma$ that is bounded
by $\gamma(\Sigma)^{O(\gamma(\Sigma))}$, and $\Gamma$ is the Gamma
function: $\Gamma(u)=\int_{0}^{\infty}t^{u-1}e^{-t}dt$.
\end{theorem}

The above result, which is critical for our analysis, has been proved
using  tools from analytic combinatorics
(see~\cite{FlajoletSedgewig:analytic-combinatorics}): singularity analysis
over expressions obtained by the symbolic method. Actually, we prefer to
translate it to the following looser form that is more convenient for our
algorithmic purposes.

\begin{corollary}
\label{corr:four} Let $\Sigma$ be a surface with boundary and let $S$ be a
set of $k$ vertices in the boundary of $\Sigma$. Let also $\gamma$ be an
integer such that $\gamma(\Sigma),\beta(\Sigma)\leq \gamma$. Then
$|\Pi_{\Sigma}(S)|=\gamma^{O(\gamma)}\cdot k^{O(\gamma)}\cdot 4^{k}$.
\end{corollary}

For every set $S$ we define ${\cal B}(S)$ as the collection of all its
partitions. Recall that if $|S|=l$, then $|{\cal B}(S)|$ is the $l$-th
Bell number and that $|{\cal B}(S)|=2^{O(l\log l)}$. Also, given a
collection ${\cal C}=\{S_{1},\ldots,S_{q}\}$ of subsets of $S$ and a
subset $S'\subseteq S$, we denote by ${\cal C}|_{S'}$ the collection of
all non-empty sets in $\{S_{1}\cap S',\ldots,S_{q}\cap S'\}$. Clearly, if
${\cal C}$ is a partition of $S$, then ${\cal C}|_{S'}$ is a partition of
$S'$.

\begin{lemma}
\label{lem:formor} Let $\Sigma$ be a surface with boundary, let $S$ be a
set of vertices in the boundary of $\Sigma$, and let $A$ be a set of
apices. Let also $\gamma$ and $k$ be integers such that
$|A|,\gamma(\Sigma),\beta(\Sigma)\leq \gamma$ and $|S|\leq k$. Then
$|\Pi_{\Sigma^{A}}(S\cup A)|=\gamma^{O(\gamma)}\cdot k^{O(\gamma)}\cdot
\gamma^{O(k)}$.
\end{lemma}

\dem Let ${\cal R}\in \Pi_{\Sigma^{A}}(S\cup A)$ and let $\mathfrak{B}$ be
an extended partition family for $(\Sigma^{A},S\cup A)$, where ${\cal
R}(\mathfrak{B})={\cal R}$. We define $\mathfrak{B}_{\Sigma}$ as the set
of connected components of the set  $(\cupall\mathfrak{B})\cap \Sigma$.
Notice that $\mathfrak{B}_{\Sigma}$ is a partition family for $(\Sigma,S)$
and thus ${\cal R}_{\Sigma}={\cal R}(\mathfrak{B}_{\Sigma})\in
\Pi_{\Sigma}(S)$. Notice also that ${\cal R}|_{A}$ is a member of ${\cal
B}(A)$. We conclude that each ${\cal R}\in \Pi_{\Sigma^{A}}(S\cup A)$
uniquely generates a pair $({\cal R}_{\Sigma},{\cal R}|_{A})\in
\Pi_{\Sigma}(S)\times {\cal B}(A)$.

We define ${\bf P}_{({\cal R}_{\Sigma},{\cal R}|_{A})}$ as the set of all
possible ${\cal R}$'s in $\Pi_{\Sigma^{A}}(S\cup A)$ that can generate a
given pair $({\cal R}_{\Sigma},{\cal R}|_{A})\in \Pi_{\Sigma}(S)\times
{\cal B}(A)$.

\begin{claim}
\label{claim:Dim1}$|{\bf P}_{({\cal R}_{\Sigma},{\cal R}|_{A})}|\leq
(|{\cal R}|_{A}|+1)^{|{\cal R}_{\Sigma}|}$.
\end{claim}

\dem We use the notation ${\cal R}|_{A}=\{A_{1},\ldots,A_{q}\}$. Let
${\cal R}\in {\bf P}_{({\cal R}_{\Sigma},{\cal R}|_{A})}$. By the above
definitions, for each $i\in\{1,\ldots,p\}$, there is a unique set, say
$P^{(i)}$, of  ${\cal R}$ containing $A_{i}$ as a subset. Moreover, there
is a (possibly empty) subset, say ${\cal B}^{(i)}$, of ${\cal R}_{\Sigma}$
such that $P^{(i)}\setminus A_{i}=\cupall {\cal B}^{(i)}$. Notice that
$\{{\cal B}^{(1)},\ldots,{\cal B}^{(i)}\}$ is a packing of ${\cal
R}_{\Sigma}$ (not necessarily a partition of ${\cal R}_{\Sigma}$, as some
sets of ${\cal R}_{\Sigma}$ may appear directly as sets in ${\cal R}$).
This means that each ${\cal R}\in {\bf P}_{({\cal R}_{\Sigma},{\cal
R}|_{A})}$ corresponds to some packing of ${\cal R}_{\Sigma}$ and some
bijection of its sets to some of the elements of ${\cal R}|_{A}$. This
corresponds to the partial functions from the set ${\cal R}_{\Sigma}$ to
the set ${\cal R}|_{A}$, that is the claimed upper bound.\findem

The rest of the proof is based on the fact that

$$|\Pi_{\Sigma^{A}}(S\cup A)|\leq \sum_{({\cal R}_{\Sigma},{\cal R}|_{A})\in\atop\Pi_{\Sigma}(S)\times {\cal B}(A)} |{\bf P}_{({\cal R}_{\Sigma},{\cal R}|_{A})}|.$$
Recall now that $|{\cal B}(A)|\leq |A|^{|A|}\leq \gamma^{\gamma}$. Also,
from Corollary~\ref{corr:four}, it holds that
$|\Pi_{\Sigma}(S)|=\gamma^{O(\gamma)}\cdot k^{O(\gamma)}\cdot 4^{k}$. The
Claim above implies that ${\bf P}_{({\cal R}_{\Sigma},{\cal R}|_{A})}\leq
(\gamma+1)^{k}$, as every packing in $\Pi_{\Sigma}(S)$ has at most
$|S|\leq k$ sets and every packing in ${\cal B}(A)$ has at most $|A|\leq
\gamma$ sets. The proof of the lemma is completed by putting all these
facts together. \findem

Let $G$ be a graph and let $S$ be a subset of $V(G)$. We define
 $\Pi_{G}(S)$ as the set of all partitions in $\Psi_{G}(S)$,
formally, $$\Pi_{G}(S)=\{{\cal R}\mid {\cal R}\in
\Psi_{G}(S)\mbox{~and~}\cupall{\cal R}=S\}.$$

%

\begin{lemma}
\label{lem:pi} Let $G$ be a graph and let $S'\subseteq S\subseteq V(G)$.
Then $|\Pi_{G}(S')|\leq |\Pi_{G}(S)|$.
\end{lemma}

\dem In order to prove the lemma, let us define an injective application
$i: \Pi_{G}(S') \hookrightarrow \Pi_{G}(S)$. Let $\mathcal{R} \in
\Pi_{G}(S')$, which implies by definition (see Section~\ref{sec:expl})
that there exists a subgraph $H \subseteq G$ whose connected components
define the packing $\mathcal{R}$ of $S'$. We define $i(\mathcal{R})$ as
the packing of $S$ given by the same subgraph $H$. It is then easy to
check that if $\mathcal{R}_1,\mathcal{R}_2 \in \Pi_{G}(S')$ with
$\mathcal{R}_1 \neq \mathcal{R}_2$, then $i(\mathcal{R}_1) \neq
i(\mathcal{R}_2)$.\findem

\begin{lemma}
\label{lemma:contr} Let $G'$ be a graph with a set $S'\subseteq V(G')$ and
an edge $e=\{x,y\}$ whose endvertices are both vertices of $S'$. Let also
$G$ be the graph obtained from $G'$ after the contraction of $e$ to a
vertex $v_{e}$, and let $S=S'\setminus \{x,y\}\cup \{v_{e}\}$. Then
$|\Pi_{G}(S)|\leq |\Pi_{G'}(S')|$.
\end{lemma}

\dem  Similarly to the proof of Lemma~\ref{lemma:contr}, let us define an
injection $i: \Pi_{G}(S) \hookrightarrow \Pi_{G'}(S')$. Let $\mathcal{R}
\in \Pi_{G}(S)$, and let $H$ be a subgraph of $G$ whose connected
components define the packing $\mathcal{R}$ of $S$. We distinguish two
cases. First, if $v_e \notin V(H)$, we define $i(\mathcal{R})$ to be the
packing of $S'$ given by the connected components of $H$. Otherwise, if
$v_e \in V(H)$, let $H' \subseteq G'$ be the graph obtained from $H$ by
removing $v_e$ and adding $x,y$, the edge $\{x,y\}$, and all the edges in
$G'$ between $x,y$ and the neighbors of $v_e$ in $H$. In this case we
define $i(\mathcal{R})$ to be the packing of $S'$ given by the connected
components of $H'$. It is again easy to check that if
$\mathcal{R}_1,\mathcal{R}_2 \in \Pi_{G}(S)$ with $\mathcal{R}_1 \neq
\mathcal{R}_2$, then $i(\mathcal{R}_1) \neq i(\mathcal{R}_2)$. \findem

The following observation gives the obvious way to enumerate packings from
partitions.

\begin{observation}
\label{obs:manyn} Let $G$ be a graph and let $S\subseteq V(G)$. Then
$\Psi_{G}(S)=\bigcup_{S'\subseteq S}\Pi_{G}(S').$
\end{observation}

Combining Lemma~\ref{lem:pi} and Observation~\ref{obs:manyn} we obtain the
following.

\begin{observation}
\label{obs:psi} Let $G$ be a graph and let $S'\subseteq S\subseteq V(G)$.
Then $|\Psi_{G}(S')|\leq |\Psi_{G}(S)|$.
\end{observation}

Let $H$ be a graph embedded in a surface $\Sigma$ with boundary. %
%
We denote by $\mathfrak{B}_{H}$ the collection of connected subsets of
$\Sigma$ corresponding to the connected components of $H$.


\begin{lemma}
\label{lem:hela} Let $G$ be a graph containing a set $A$ of vertices such
that $G\setminus A$  is embedded in a surface $\Sigma$. Let also $S$ be
the set of vertices of $G$ that lie on the boundary of $\Sigma$. Then
$|\Pi_{G}(S\cup A)|\leq |\Pi_{\Sigma^{A}}(S\cup A)|$.
\end{lemma}

\dem It is enough to prove that for every partition ${\cal R}$ in
$\Pi_{G}(S\cup A)$ there is an extended partition family $\mathfrak{B}$
for $(\Sigma^{A},S\cup A)$ such that ${\cal R}(\mathfrak{B})={\cal R}$.
For this, consider a subgraph $H$ of $G$ where ${\cal P}_{S\cup
A}(H)={\cal R}$. As ${\cal R}\in \Pi_{G}(S\cup A)$, it holds that $\cupall
{\cal R}=S\cup A$ and therefore $\cupall {\cal R}\subseteq V(H)$. As
$H\setminus A$ can be embedded in $\Sigma$, the set
$\mathfrak{B}_{H\setminus A}$ is a partition family for $(\Sigma,S)$. Let
now $H_{A}$ be the subgraph of $H$ formed by its edges that are not
embedded in $\Sigma$. Observe that $H_{A}$ is isomorphic to a subgraph of
$C_{A}$ and therefore its edges can be seen as a collection
$\mathfrak{E}_{A}$ of flying edges where each apex vertex is contained in
some edge of $\mathfrak{E}_{A}$. Let $\mathfrak{B}$
be the connected components of the set $\cupall (%
\mathfrak{B}_{H\setminus A}\cup
\mathfrak{E}_{A}%
)$. Clearly, $\mathfrak{B}$ is an extended partition family for
$(\Sigma^{A},S\cup A)$. It is now easy to verify that ${\cal
R}(\mathfrak{B})={\cal R}$ and the lemma follows. \findem

\begin{lemma}
\label{lem:hel} Let  $G$ be a graph containing a set $A$ of vertices such
that $G\setminus A$  is embedded in a surface $\Sigma$ with boundary. Let
also $S$ be the set of vertices of $G$ that lie on the boundary of
$\Sigma$ and $A'\subseteq A$. Then, if $|S|\leq k$ and
$|A|,\gamma(\Sigma),\beta(\Sigma)\leq \gamma$, then $|\Psi_{G}(S\cup A')|=
\gamma^{O(\gamma)}\cdot k^{O(\gamma)}\cdot \gamma^{O(k)}$.
\end{lemma}

\dem From Observation~\ref{obs:psi}, it is enough to prove the lemma for
the case where $A'=A$.
From Lemmata~\ref{lem:formor} and~\ref{lem:hela}, it follows that
$|\Pi_{G}(S\cup A)|=\gamma^{O(\gamma)}\cdot k^{O(\gamma)}\cdot
\gamma^{O(k)}$. From Lemma~\ref{lem:pi}, we obtain that $|\Pi_{G}(W)|\leq
|\Pi_{G}(S\cup A)|=\gamma^{O(\gamma)}\cdot k^{O(\gamma)}\cdot
\gamma^{O(k)}$ for every $W\subseteq S\cup A$. Therefore, from
Observation~\ref{obs:manyn}, $|\Psi_{G}(S\cup A)|\leq 2^{|S|+|A|}\cdot
\gamma^{O(\gamma)}\cdot k^{O(\gamma)}\cdot
\gamma^{O(k)}=\gamma^{O(\gamma)}\cdot k^{O(\gamma)}\cdot \gamma^{O(k)}$
and the lemma follows. \findem


Let $\Sigma$ be a surface without boundary, and let $\mathcal{N}$ be a set
of $O$-arcs in $\Sigma$ pairwise intersecting at zero-dimensional subsets
of $\Sigma$. Then the closure of each connected component of $\Sigma
\setminus \cupall\mathcal{N}$ is called a \emph{pseudo-surface}. Notice
that the boundary of a pseudo-surface is a subset of $\mathcal{N}$ and
that the definition of the parameter $\theta(\mathcal{N})$ introduced in
Section~\ref{sec:topological_lemmas} can be naturally extended to
pseudo-surfaces.
 If $\Sigma$ is a pseudo-surface with boundary given by a
finite set $\mathcal{N}$ of $O$-arcs pairwise intersecting at finite
zero-dimensional subsets of $\Sigma$, note that $\Sigma$ is a surface with
boundary if and only if $\theta(\mathcal{N})=0$. Note also that the
closure of each of the two connected components in the last condition of
Definition~\ref{def:surface_cut} is a pseudo-surface.


\begin{lemma}
\label{lem:pseu} Let $G$ be a graph embedded in a pseudo-surface $\Sigma$
whose boundary is given by a collection $\mathcal{N}$ of nooses of $G$
pairwise intersecting only at vertices of $G$, and such that
$\theta(\mathcal{N}) > 0$. Let $S$ be the set of vertices of $G$ that lie
on the boundary of $\Sigma$. Then there is a graph $G'$ embedded in a
pseudo-surface $\Sigma'$ with boundary given by a collection
$\mathcal{N}'$ of nooses of $G'$, such that
\begin{itemize}
\item  $\theta(\mathcal{N}')= \theta(\mathcal{N})-1$;\vspace{-0.12cm}
\item  $G$ is the result of the contraction of  an edge in $G'$;\vspace{-0.12cm}
\item if $S'$ is the set of vertices of $G'$ that lie on the boundary of $\Sigma'$, then $|S'|=|S|+1$.
\end{itemize}
\end{lemma}

\begin{figure}[t]
\begin{center}
\vspace{-0.3cm}
\includegraphics[width=14.5cm]{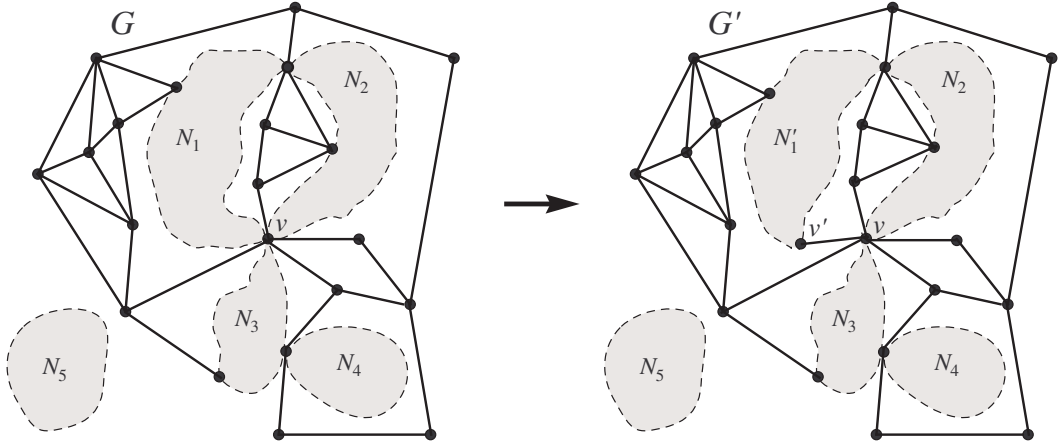}
\end{center}
\vspace{-0.5cm} \caption{Example of the construction of $\Sigma'$ and $G'$
in the proof of Lemma~\ref{lem:pseu}. On the left, we have a graph $G$
(depicted with thick lines) embedded in a pseudo-surface $\Sigma$ whose
boundary is given by the set of nooses
$\mathcal{N}=\{N_1,N_2,N_3,N_4,N_5\}$ (in grey) pairwise intersecting at
vertices of $G$, with $\theta(\mathcal{N})=4$. On the right, the
corresponding graph $G'$ embedded in a pseudo-surface $\Sigma'$ with
boundary given by $\mathcal{N}'=\{N_1',N_2,N_3,N_4,N_5\}$, and such that
$\theta(\mathcal{N}')=3$. In this example, we have that $|S|=6$ and
$|S'|=7$.}\label{fig:InverseContraction}
\end{figure}

\dem Without loss of generality, let $v \in N_1 \cap \ldots \cap
N_{\ell}$, with $N_1,\ldots,N_{\ell} \in \mathcal{N}$ and $\ell \geq 2$,
so by assumption $v \in S \subseteq V(G)$; for an illustration throughout
the proof, see Figure~\ref{fig:InverseContraction}. We build from $\Sigma$
a pseudo-surface $\Sigma'$ by replacing noose $N_1$ with a noose $N_1'$
obtained from $N_1$ by slightly deforming it around $v$ in such a way that
$v \notin N_1'$ (note that this is clearly possible, as by assumption the
nooses intersect only at vertices of $G$). As the nooses in $\Sigma$ and
in $\Sigma'$ intersect at the same vertices except for vertex $v$, we have
that $\theta(\mathcal{N}')= \theta(\mathcal{N})-1$. We now construct $G'$
from $G$ as follows: We start from the embedding of $G$ in $\Sigma$, and
we embed it in $\Sigma'$ in such a way that $v \in N_2 \cap \ldots \cap
N_{\ell}$. Finally, we add a new vertex $v' \in N_1'$ and we add the edge
$\{v,v'\}$. By construction, it is clear that $G$ can be obtained from
$G'$ by contracting edge $\{v,v'\}$, and that $S'=S \cup \{v'\}$. \findem

\bigskip

\noindent{\bf Proof of Lemma~\ref{lem:jbf}:} In case $|\mids(e)\setminus
A|\leq 2$, we have that $|\mids(e)|=O(\gamma)$ and the result follows as
$|\Pi_{G_{e}}(\mids(e))|\leq |{\bf B}(O(\gamma))|= 2^{O(\gamma\log
\gamma)}$. In the remaining case, let $H$ be the graph of the polyhedral
decomposition $({\cal G},A)$ of $G$ that corresponds to edge $e$. Let also
${\cal N}$ be the corresponding set of $O(\gamma)$ nooses meeting all
vertices of $\mids(e)\setminus A$. Let  also $\Sigma^{*}$ be the closure
of the connected component of $\Sigma\setminus\bigcup_{N\in{\cal N}}N$
where the graph $G_{e}\setminus A$ is embedded. Clearly, $\Sigma^{*}$ is a
pseudo-surface with boundary given by a set of nooses $\mathcal{N}$ with
$\theta(\mathcal{N})=O(\gamma)$. By inductively applying
Lemmata~\ref{lemma:contr} and~\ref{lem:pseu}, we can assume that
$\Sigma^{*}$ is a surface with boundary such that
$O(|\mids(e)|+\gamma(\Sigma))=O(k+\gamma)$ of the vertices of $G_{e}$ lie
on this boundary. Then the result follows directly from
Lemma~\ref{lem:hel} by setting $G_{e}$ instead of $G$, $\Sigma^{*}$
instead of $\Sigma$,  $A \cap \mids(e)$ instead of $A'$, and $A \cap
V(G_e)$ instead of $A$. \findem

\section{Conclusions and open problems}
\label{sec:conclusions}


As stated in Theorem~\ref{teo:finals}, our results can be summarized as
follows:  Every connected packing-encodable problem whose input graph $G$
is embedded in a surface of Euler genus $\gamma$, and has branchwidth at
most $k$, can be solved in $\gamma^{O(k)} \cdot k^{O(\gamma)} \cdot
\gamma^{O(\gamma)}\cdot n^{O(1)}$ steps.

As we mentioned, the problems tackled in~\cite{DFT06} can be encoded with
pairings, and therefore they can be seen as special cases of
packing-encodable problems. As a result of this, we reproduce all the
results of~\cite{DFT06}. Moreover, as our approach does not use
planarization, our analysis provides algorithms where the dependence on
the Euler genus $\gamma$ is better than the one in~\cite{DFT06}. In
particular, the running time of the algorithms in~\cite{DFT06} is
$2^{O(\gamma\cdot k+\gamma^{2}\cdot \log k)}\cdot n^{O(1)}$, while in our
case the running time is $2^{O(\log \gamma \cdot k + \gamma \cdot \log k +
\gamma \cdot \log \gamma)}\cdot n^{O(1)}$.

Dynamic programming is important for the design of {\em subexponential}
exact or parameterized algorithms. Using the fact that bounded-genus
graphs have branchwidth at most $O(\sqrt{\gamma \cdot n})$~\cite{FoTh04},
we derive the existence of exact algorithms in $O^{*}(2^{O(\log \gamma
\cdot \sqrt{\gamma n}+\gamma \cdot \log n+ \gamma \cdot \log \gamma})$
steps for all connected packing-encodable problems. Moreover, using
bidimensionality theory (see~\cite{DFHT05,DHT06}), one can derive
$2^{O(\gamma \cdot \log \gamma \cdot \sqrt{k} + \gamma \cdot \log k)}\cdot
n^{O(1)}$ time parameterized algorithms for all bidimensional connected
packing-encodable problems, where here $k$ is the corresponding parameter.



Note that the running time of our algorithms is conditioned by the
construction of an appropriate surface cut decomposition. This
preprocessing step takes $2^{3k+O(\log k)} \cdot n^3$ steps by
Theorem~\ref{teo:surface_cut}. Finding a preprocessing algorithm with
better polynomial dependance remains open. As finding an optimal branch
decomposition of a surface-embedded graph in polynomial time is open, it
may be even possible that computing an optimal surface cut decomposition
can be done in polynomial time.

Sometimes dynamic programming demands even more complicated encodings. We
believe that our results can also serve in this direction. For instance,
surface cut decompositions have recently been used in~\cite{ADF+10} for
minor containment problems, where tables encode partitions of packings of
the middle sets.

A natural extension of our results is to consider more general classes of
graphs than bounded-genus graphs. This has been done in~\cite{DFT08} for
problems  where the tables of the algorithms encode pairings of the middle
sets. Extending these results for connected packing-encodable problems
(where tables encode subsets of the middle sets) using the planarization
approach of~\cite{DFT08} appears to be a quite complicated task. We
believe that our surface-oriented approach could be more successful in
this direction and we find it an interesting,
but non-trivial task~\cite{RST11minors}.\\ 




\noindent \textbf{Acknowledgement}. We would like to thank Sergio Cabello
for inspiring discussions and for
 pointing us to several helpful topological references.

{\bibliography{algo}}
\bibliographystyle{abbrv}
\include{bib}

\newpage

\section*{Appendix}

\subsection*{Two examples of dynamic programming algorithms}

In this Appendix we present two examples of typical dynamic programming
algorithms on graphs of bounded branchwidth. The first algorithm solves
the {\sc Vertex Cover} problem, which is a problem whose solutions can be
simply encoded by a subset of vertices.  The second algorithm solves the
{\sc Connected Vertex Cover} problem, which is a packing-encodable
problem, but cannot be encoded by neither a subset nor a pairing of
vertices.


%

\paragraph{Dynamic programming for {\sc Vertex Cover}.} Given a graph $G$ and a non-negative  integer $\ell$, we have to decide
whether $G$ contains a set $S\subseteq V(G), |S|\leq \ell$, meeting all
edges of $G$.


Let $G$ be a graph and $X,X'\subseteq V(G)$ where $X\cap X'=\emptyset$. We
say that $\vc(G,X,X')\leq \ell$ if $G$ contains a vertex cover $S$ where
$|S|\leq \ell$ and $X\subseteq S\subseteq V(G)\setminus X'$.
%
Let ${\cal R}_{e}=\{(X,\ell)\mid X\subseteq \mids(e) \mbox{~and~}
\vc(G_{e},X,\mids(e)\setminus X)\leq \ell\}$  and observe that $\vc(G)\leq
\ell$ if and only if $(\emptyset,\ell)\in {\cal R}_{e_{r}}$.
 For each $e\in E(T)$ we can compute ${\cal R}_{e}$ by using the following dynamic programming formula:
 \begin{eqnarray*}
 {\cal R}_{e} & = &
\begin{cases}
\{(X,\ell)\mid X\subseteq e \mbox{~and~} X\neq \emptyset \wedge \ell\geq |X|\}  & \text{if $e\in L(T)$}\\
\{(X,\ell)\mid \exists (X_{1},\ell_{1})\in {\cal R}_{e_{1}},  \exists (X_{2},\ell_{2})\in {\cal R}_{e_{2}}:   & \\
(X_{1}\cup X_{2})\cap \mids(e)=X \wedge \ell_{1}+\ell_{2}-|X_{1}\cap
X_{2}|\leq \ell \} & \text{if $e\not\in L(T)$}
\end{cases}
\end{eqnarray*}
Note that for each $e\in E(T)$, $|{\cal R}_{e}|\leq 2^{|\mids(e)|}\cdot
\ell$. Therefore, the above algorithm can check whether $\vc(G)\leq \ell$
in $O(4^{\bw(G)}\cdot \ell^2\cdot |V(T)|)$ steps. Clearly, this simple
algorithm is single-exponential in $\bw(G)$. Moreover the above dynamic
programming machinery can be adapted to many other combinatorial problems
where the certificate of the solution is a (non-restricted) subset of
vertices (e.g. {\sc Dominating Set}, {\sc 3-Coloring},
{\sc Independent Set}, among others). 

\paragraph{Dynamic programming for {\sc Connected Vertex Cover}.}
Suppose now that we are looking for a {\em connected} vertex cover of size
$\leq \ell$.  Clearly, the above dynamic programming formula does not work
for this variant as we should keep track of more information on $X$
towards encoding the connectivity demand.

Let $G$ be a graph, $X\subseteq V(G)$ and ${\cal H}$ be a (possibly empty)
hypergraph whose vertex set is a subset of $X$, whose  hyperedges are
non-empty, pairwise non-intersecting, and such that each vertex of ${\cal
H}$ belongs to some of its hyperedges (we call such a hypergraph {\em
partial packing} of $X$).
Suppose that ${\cal H}$ is a partial packing on $\mids(e)$. We say that
$\cvc(G,{\cal H})\leq \ell$ if $G$ contains a vertex cover $S$ where
$|S|\leq \ell$ and such that if ${\cal C}$ is the collection of the
connected components of $G_{e}[S]$, then either $|E({\cal H})|=|{\cal C}|$
and $(X,\{X\cap V(C)\mid C\in {\cal C}\})={\cal H}$ or
 $E({\cal H})=\emptyset$ and $|{\cal C}|=1$.

As before, let ${\cal Q}_{e}=\{({\cal H},\ell)\mid \cvc(G,{\cal H})\leq
\ell\}$ and observe that $\cvc(G)\leq  \ell$ if and only if
$((\emptyset,\emptyset),\ell)\in {\cal Q}_{e_{r}}$. The dynamic
programming formula for computing ${\cal Q}_{e}$ for each $e\in E(T)$ is
the following.
 \begin{eqnarray*}
 {\cal Q}_{e} & = &
\begin{cases}
\{({\cal H},\ell)\mid  \min\{\ell, |E({\cal H})|+1 \} \geq |V({\cal H})| \geq 1  & \text{if $e\in L(T)$}\\
\{({\cal H},\ell)\mid \exists ({\cal H}_{1},\ell_{1})\in {\cal Q}_{e_{1}},  \exists ({\cal H}_{2},\ell_{2})\in {\cal Q}_{e_{2}}:   & \\
V({\cal H}_{1})\cap (\mids(e_1) \cap \mids(e_2)) = V({\cal H}_{2})\cap (\mids(e_1) \cap \mids(e_2)), & \\
({\cal H}_{1}\oplus {\cal H}_{2})[\mids(e)]={\cal H}, \ell_{1}+\ell_{2}-|V({\cal H}_{1})\cap V({\cal H}_{2})|\leq \ell  \}, & \\
\mbox{~if~}E({\cal H})=\emptyset\mbox{~then~} |E({\cal H}_{1}\oplus {\cal H}_{2})|=1, \mbox{and}\\
\mbox{~if~}E({\cal H}) \neq\emptyset\mbox{~then~} |E(H_{1}\oplus
H_{2})|=|E({\cal H})| &  \text{if $e\not\in L(T)$}.\
\\
\end{cases}
\end{eqnarray*}
In the above formula,  ${\cal H}_{1}\oplus{\cal H}_{2}$ is the hypergraph
with vertex set $V({\cal H}_{1})\cup V({\cal H}_{2})$ where  each of its
hyperedges contains the vertices of each of the connected components of
${\cal H}_{1}\cup {\cal H}_{2}$.

Clearly, each ${\cal H}$ corresponds to a collection of subsets of $X$ and
the number of such collections for a given set $\mids(e)$ of $r$ elements
is given by the $r$-th Bell number of $r$, denoted by $B_r$.  By taking
the straightforward upper bound $|B_{r}|=2^{O(r\log r)}$, we have that one
can check whether an input graph $G$ has a connected vertex cover of size
at most $\ell$ in $2^{O(\bw(G)\cdot \log \bw(G))}\cdot \ell \cdot |V(T)|$
steps.

As the growth of $B_{r}$ is not single-exponential, we cannot hope for a
single-exponential (in $\bw(G)$) running time for the above dynamic
programming procedure, and no deterministic algorithm is known for this
problem running in time single-exponential in $\bw(G)$. The same problem
appears for numerous other problems where further restrictions apply to
their solution certificates. Such problems can be connected variants of
problems encodable by a subset of vertices, and others such as {\sc
Maximum Induced Forest}, \textsc{Maximum $d$-Degree-Bounded Connected
Subgraph}, \textsc{Metric TSP}, \textsc{Maximum $d$-Degree-Bounded
Connected Induced Subgraph} and all the variants studied in~\cite{SaTh10},
\textsc{Connected Dominating Set}, \textsc{Connected $r$-Domination}, {\sc
Feedback Vertex Set}, \textsc{Connected Feedback Vertex Set},
\textsc{Maximum Leaf Spanning Tree}, \textsc{Maximum Full-Degree Spanning Tree}, 
\textsc{Steiner Tree}, or \textsc{Maximum Leaf Tree}.

%

%

%


%


\end{document}